\pdfoutput=1
\documentclass[journal]{IEEEtran}

\usepackage{etoolbox}

\usepackage[dvipsnames]{xcolor}
\usepackage{graphicx}
\usepackage{enumitem}
\usepackage{tabularx}
\usepackage{array}

\usepackage{booktabs}
\newcolumntype{V}{>{\centering\arraybackslash} m{0.16\textwidth} }
\newcolumntype{v}{>{\centering\arraybackslash} m{6pt} }

\usepackage{multirow}

\usepackage{makecell}  

\usepackage[caption=false,font=normalsize,labelfont=sf,textfont=sf]{subfig}

\usepackage{tikz}
\usepackage{adjustbox}
\usetikzlibrary{3d, calc}
\usetikzlibrary{patterns}

\usepackage{hyperref}
\hypersetup{
  colorlinks=true,
  linkcolor=red,
  filecolor=magenta,
  urlcolor=cyan,
  citecolor=blue,
}

\definecolor{darkgreen}{cmyk}{0.8,0,0.8,0.45}
\definecolor{lightgreen}{cmyk}{0.8,0,0.8,0.25}
\usepackage[ruled,vlined,linesnumbered]{algorithm2e}

\SetCommentSty{mycommfont}
\SetKwInput{Parameter}{Parameters} 

\let\oldnl\nl
\newcommand{\nonl}{\renewcommand{\nl}{\let\nl\oldnl}} 

\DeclareGraphicsExtensions{.pdf,.jpg,.png,.eps}
\graphicspath{{figs/}}

\usepackage{mdframed}
\newmdenv[
  linecolor=red,
  backgroundcolor=red!10,
  linewidth=2pt,
  roundcorner=5pt,
  font=\large\bfseries,
]{warning}

\usepackage{amsmath,amsfonts}
\usepackage{amssymb}
\usepackage{mathtools}

\usepackage{dsfont}
\usepackage{stmaryrd}
\SetSymbolFont{stmry}{bold}{U}{stmry}{m}{n}
\usepackage{bm}  

\newcommand{\widebar}{\overline}

\newcommand{\blockind}{b}

\newcommand{\cst}[1]{{#1}}
\newcommand{\vect}[1]{{\bm{#1}}}

\newcommand{\ndop}[1]{\mathcal{#1}}

\newcommand{\upd}[2]{{#1}^{(#2)}}

\DeclareMathAlphabet\mathbfcal{OMS}{cmsy}{b}{n}

\newcommand{\x}{\bm{x}}
\newcommand{\y}{\bm{y}}
\newcommand{\z}{\bm{z}}

\newcommand{\uv}{\bm{u}}

\newcommand{\w}{\bm{w}}

\newcommand{\Hm}{\bm{H}}

\newcommand{\N}{\cst{N}}
\newcommand{\M}{\cst{M}}

\newcommand{\nblocks}{B}
\newcommand{\nlayers}{K}

\newcommand{\BS}{\mathbb{V}}
\newcommand{\IS}{\mathbb{I}}

\newcommand{\block}[1]{{#1}_b}
\newcommand{\blockp}[1]{\block{\left[#1\right]}}

\newcommand{\xb}{\block{\x}}
\newcommand{\yb}{\block{\y}}
\newcommand{\zb}{\block{\z}}

\newcommand{\Nb}{{\block{\N}}}

\newcommand{\Sb}{\block{\bm{S}}}

\newcommand{\R}[1]{\mathbb{R}^{#1}}

\DeclarePairedDelimiter{\floor}{\lfloor}{\rfloor}

\DeclareMathOperator*{\argmin}{argmin}
\DeclareMathOperator{\prox}{prox}
\DeclareMathOperator{\proj}{proj}
\DeclareMathOperator{\card}{\sharp}

\DeclareMathOperator{\KL}{D_{KL}}

\DeclareMathOperator{\snr}{SNR}

\usepackage{theorem}
\usepackage{thmtools}

\theoremstyle{plain}{\theorembodyfont{\rmfamily}%
}
\theoremstyle{plain}{\theorembodyfont{\rmfamily}%
\newtheorem{assumption}{Assumption}}
\theoremstyle{plain}{\theorembodyfont{\rmfamily}%
}
\theoremstyle{plain}{\theorembodyfont{\rmfamily}%
}
\theoremstyle{plain}{\theorembodyfont{\rmfamily}%
\newtheorem{example}{Example}}
\theoremstyle{plain}{\theorembodyfont{\rmfamily}%
\newtheorem{remark}{Remark}}
\theoremstyle{plain}{\theorembodyfont{\rmfamily}%
\newtheorem{definition}{Definition}}

\usepackage{cleveref}
\crefname{definition}{definition}{definition}
\Crefname{definition}{Definition}{Definitions}
\crefname{property}{property}{property}
\Crefname{property}{Property}{Properties}
\crefname{assumption}{assumption}{assumptions}
\Crefname{assumption}{Assumption}{Assumptions}
\crefname{remark}{remark}{remarks}
\Crefname{remark}{Remark}{Remarks}
\crefname{algocf}{algorithm}{algorithms}
\Crefname{algocf}{Algorithm}{Algorithms}

\usepackage{outlines}
\usepackage[normalem]{ulem}  

\NewDocumentEnvironment{blueenv}{}{\color{Blue}}{}

\NewDocumentEnvironment{greenenv}{}{\color{OliveGreen}}{}

\NewDocumentCommand{\mb}{m}{\textcolor{black}{#1}}

\begin{document}


\title{A Distributed Plug-and-Play MCMC Algorithm \\ for High-Dimensional Inverse Problems}
\author{Maxime Bouton, Pierre-Antoine Thouvenin, Audrey Repetti and Pierre Chainais%
\thanks{This work was supported by the ANR project ``Chaire IA Sherlock'' ANR-20-CHIA-0031-01 hold by P. Chainais, the national support within the {\em programme d'investissements d'avenir} ANR-16-IDEX-0004 ULNE, Région HDF, and the CNRS IEA ``DAISHI'' hold by P.-A. Thouvenin and A. Repetti. 
The work of A. Repetti was partly supported by the EPSRC grant EP/X028860 and a grant from the Simons Foundation within the Isaac Newton Programme.
HPC and storage resources were provided by GENCI at IDRIS on the supercomputer Jean Zay's V100 partition \emph{via} the grant 2024-AD010615597.}%
}



\maketitle

\begin{abstract} 
%
Markov Chain Monte Carlo (MCMC) algorithms are standard approaches
to solve imaging inverse problems and quantify estimation uncertainties, a key requirement in absence of ground-truth data.
To improve estimation quality, Plug-and-Play MCMC algorithms, such as PnP-ULA, have been recently developed to accommodate priors encoded by a denoising neural network. 
Designing scalable samplers for high-dimensional imaging inverse problems remains a challenge: drawing and storing high-dimensional samples can be prohibitive, especially for high-resolution images.
To address this issue, this work proposes a distributed sampler based on approximate data augmentation and PnP-ULA to solve very large problems.
The proposed sampler uses lightweight denoising convolutional neural network, to efficiently exploit multiple GPUs on a Single Program Multiple Data architecture.
Reconstruction performance and scalability are evaluated on several imaging problems. Communication and computation overheads due to the denoiser are carefully discussed.
The proposed distributed approach noticeably combines three very precious qualities: it is scalable, enables uncertainty quantification, for a reconstruction performance comparable to other PnP methods.
\end{abstract}


\section{Introduction}
\label{sec:intro}

%
This work focuses on high-dimensional imaging inverse problems, aimed at recovering an unknown image $\overline{\bm{x}} \in \mathbb{R}^{N}$ from degraded observations $\bm{y} \in \mathbb{R}^{M}$, linked by a model
\begin{equation}
    \label{eq:forward_model}
    \y = \ndop{A}(\overline{\x}),
\end{equation}
where $\ndop{A}: \R{\N} \to \R{\M}$ models both the deterministic degradations and the noise affecting $\overline{\bm{x}}$. 
Both $\M$ and $\N$ can range from $10^6$ to $10^{10}$, raising computational challenges related to prohibitive memory requirements and runtime.
Bayesian inference is a usual approach to account for uncertainties, based on the posterior distribution combining the likelihood associated with~\eqref{eq:forward_model} with prior information on $\overline{\bm{x}}$~\cite{Robert2007}.
In this paper, we consider problems whose posterior distribution is described by a probability density function (pdf) of the form
\begin{equation} 
    \label{eq:generic_posterior_distribution}
    \pi(\x|\y) 
    \propto 
    p(\x) \, \exp\big(- f_1 (\Hm_1 \x) - f_2 (\Hm_2 \x) \big),
\end{equation}
where $f_1 \colon \R{\M_1} \to (-\infty, +\infty ]$ is a proper, Lipschitz-differentiable function, $\Hm_1 \colon \R{\N} \to \R{\M_1}$ and $\Hm_2\colon \R{\N}\to \R{\M_2}$ are linear operators, and $f_2 \colon \R{\M_2} \to (-\infty, +\infty ]$ is a proper, lower semi-continuous (l.s.c) convex function.
The function $p\colon \R{\N} \to [0, 1]$ can represent a pdf corresponding to part of the prior distribution, based for instance on a neural network (NN)~\cite{Holden2022,Laumont2022,Faye2025}.
Conditions to ensure the existence of a proper prior with density $p$ have for instance been investigated in~\cite{Laumont2022} for a denoising NN.
The dependence on $\y$ is omitted in~\eqref{eq:generic_posterior_distribution}, so that $f_1 \circ \Hm_1 $ and $f_2 \circ \Hm_2$ can come from either the likelihood or the prior distributions.

%
Solving~\eqref{eq:forward_model} in a high-dimensional setting is challenging, especially when quantifying uncertainties, which is critical for decision-making processes, 
and when calibrated data is hard or impossible to obtain (\emph{e.g.}, in astronomy~\cite{Liaudat2024}).
%
Markov-chain Monte Carlo (MCMC) algorithms are commonly used to explore the posterior distribution~\eqref{eq:generic_posterior_distribution}, allowing estimates to be formed with credibility intervals to quantify uncertainties.
Nevertheless, they usually do not scale well.
%
%
%
Different approaches have been proposed in the recent years to design scalable MCMC algorithms.
For instance, samplers inspired by primal-dual proximal algorithms~\cite{Komodakis2015} have been proposed in~\cite{Lau2024,Burger2024}.
An approximate data augmentation approach similar to half-quadratic splitting strategy~\cite{Geman1995} has been developed in \cite{Vono2020}, addressed with a Gibbs samplers referred to as Split Gibbs sampler (SGS).
A distributed sampler based on \cite{Vono2020} has been proposed in \cite{Thouvenin2024jcgs}, exploiting the hypergraph structure underlying the approximate posterior distribution~\eqref{eq:generic_posterior_distribution} to partition both $\y$ and $\x$ over several workers.
This enables Proximal Stochastic Gradient Langevin Algorithms (PSGLA)~\cite{Salim2019} transitions to be implemented within SGS on a single program multiple data (SPMD) architecture~\cite{Darema2001}.
In this setting, each worker conducts the same operations on disjoint subsets of $\y$ and $\x$, with a small number of communications whenever the operations involved are localized.
Compared to a client-server implementation, this approach was shown to offer more flexibility to (i) use multiple workers with a balanced computing load, and (ii) limit communication costs~\cite{Thouvenin2024jcgs}.

%
Finally, MCMC methods have recently been paired with data-driven priors through the use of NNs to achieve high-quality estimation.
Learned priors encoded by NNs have emerged as powerful tools to incorporate data-driven knowledge~\cite{Ongie2020}.
%
%
In particular, Plug-and-Play (PnP) approaches~\cite{Venkatakrishnan2013,Zhang2021} leverage a generic denoiser within a standard inference algorithm.
Several PnP MCMC algorithms have been proposed using different families of learned priors~\cite{Laumont2022,Cai2024,Coeurdoux2024pnp,Wu2024,Faye2025}.
However, none of these algorithms has been designed with a strong focus on scalability and distributed computing.

This work is aimed at designing a distributed data-driven MCMC algorithm to achieve state-of-the-art reconstruction quality, while enabling uncertainty quantification in high dimensional settings.
\mb{Our main contribution is to propose a multi-GPU distributed MCMC algorithm to address problems, whose size exceeds the memory of a single GPU. The proposed algorithm relies on the structure of a recent distributed Gibbs sampler~\cite{Thouvenin2024jcgs} designed for imaging inverse problems, which is statistically equivalent to its serial counterpart. Specifically, our approach relies on the design of a distributed framework for the NN encoding the prior used in the sampler.} 
In this context, we focus on fairly light NN architectures, as the number of parameters of the NN can have a notable memory cost for distributed computing.

%
The main classes of denoisers used in PnP-MCMC algorithms are convolutional NN (CNN)~\cite{Fleuret2024}, Normalizing Flows (NF)~\cite{Kobyzev2021} and Denoising Diffusion Models (DDM)~\cite{Daras2024}.
Both NF and DDM are usually encoded by a very large number of parameters and involve non-local operations, which precludes an efficient distributed implementation.
In contrast, CNNs offer promising opportunities for distributed computing.
Each of their layers only involve convolution operators and, in general, entry-wise non-linearities. 
The former can usually be distributed efficiently, whereas the latter are embarrassingly parallel.
Besides, CNNs designed using unrolled algorithms~\cite{Monga2021}, such as the Deep Dual Forward-Backward (DDFB) network~\cite{Repetti2022eusipco,Le2024}, 
have been shown to offer a good compromise between restoration performance and reduction in the number of parameters compared to other alternatives.

%
Inspired by~\cite{Vono2020, Thouvenin2024jcgs}, we propose a distributed sampler based on the PnP unadjusted Langevin Algorithm (PnP-ULA)~\cite{Laumont2022}, using CNNs as pre-trained denoisers.
In a preliminary work~\cite{Bouton2025ssp}, we presented simulation results suggesting that a distributed version of PnP-ULA based on DDFB allows for significant scalability and acceleration gains.
In the current work, we significantly extend these results to general CNNs, describing (i) the distributed approach adopted for all the operators involved in the proposed sampler (including the CNNs), and (ii) the computation and communication overheads induced by different choices of CNN denoisers, namely DnCNN~\cite{Zhang2017}, DRUNet~\cite{Zhang2021} and DDFB.
We also provide extensive simulation results to validate the scalability and reconstruction quality of the proposed sampler.

%
The paper is organized as follows.
\Cref{sec:methodology} formulates the problem and introduces the proposed sampler. 
Locality assumptions for an efficient distributed implementation are introduced in~\cref{sec:prob_statement}.
\Cref{sec:denoiser} discusses denoiser structures favorable to a distributed computing setting, and describes the proposed distributed implementation on an SPMD architecture.
\Cref{sec:dsgs_pnp_ula_psgla} exploits the localized structure of the denoiser to design the proposed distributed PnP-MCMC algorithm.
\Cref{sec:experiments} presents the experiment settings used to evaluate the proposed algorithm in terms of reconstruction quality and scalability.
Scalability potential is discussed for different representative choices of denoisers in terms of memory costs, communication costs and runtime.
Experiment results are reported and discussed in~\Cref{sec:exp_results}.
\Cref{sec:conclusions} gathers conclusions and research perspectives.

\smallskip\noindent\textbf{Notation. ~}
The union and disjoint union of sets are denoted by $\cup$ and $\sqcup$, respectively.
The cardinality of a set is denoted by $\card $.
The adjoint of $\bm{H} \in \mathbb{R}^{M \times N}$ is denoted by $\bm{H}^*$, and $\mathbf{1}_N \in \mathbb{R}^N$ is the unit vector with all coefficients equal to 1. 
The indicator function of a non-empty set $\mathcal{C} \subset \mathbb{R}^N$ is denoted by $\iota_{\mathcal{C}}$, with $\iota_{\mathcal{C}}(\bm{x}) = 0$ if $\bm{x} \in \mathcal{C}$, and $+\infty$ otherwise, and we further use the notation $\mathds{1}_\mathcal{C}: \x \mapsto e^{-\iota_\mathcal{C} (\x)}$.
The proximity operator~\cite{Moreau1965} of $\varphi\colon \mathbb{R}^N \to (-\infty,+\infty]$ \mb{at $\x = (x^{(n)})_{1\leq n\leq N} \in \mathbb{R}^N$} is defined as $\prox_{\varphi}(\x) = \argmin_{\uv \in \mathbb{R}^N} \left\{\varphi(\uv) + \frac{1}{2}\|\uv - \x\|^2_2\right\}$.
In particular, $\prox_{\iota_{\mathcal{C}}}$ is \mb{the orthogonal projector $\proj_{\mathcal{C}}$ onto $\mathcal{C}$.}


\section{Proposed approach}
\label{sec:methodology}

This section describes the proposed approach, building on SGS~\cite{Vono2020} and PnP-ULA~\cite{Laumont2022}.
\Cref{sec:prob_statement} introduces the model considered, specifying the operator locality assumptions required for an efficient distributed sampling strategy.
\Cref{sec:denoiser} discusses several CNNs amenable to an efficient distributed implementation.
\Cref{sec:dsgs_pnp_ula_psgla} finally introduces the proposed distributed sampler.

\subsection{Problem statement} \label{sec:prob_statement}

\noindent\textbf{Operator locality.~}
We consider the case when operators in~\eqref{eq:generic_posterior_distribution} only act locally with respect to a partition of $\{1, \dotsc, N\}$.
Then, the algorithm can be split into loosely-dependent tasks, that can be carried out by different workers. 
Dependencies between the tasks are handled by lightweight communications between the workers.
Operator locality is formalized below.

\begin{definition}[Local selection operators] \label{prop:localselection}   
    Let $(\BS_b)_{1\leq \blockind\leq B}$ be a partition of $\{1, \ldots, N\}$ into $B>1$ subsets, and, for every $b \in \{1, \ldots, B\}$, $\card \BS_b = N_b \neq 0$. Let $(\widetilde{\BS}_b)_{1\leq \blockind\leq B} $ be an extension of this partition, such that, for every $b \in \{1, \ldots, B\}$, 
    $\BS_b \subseteq \widetilde{\BS}_b$ and $\card \widetilde{\BS}_b = \widetilde{N}_b > N_b$.
    Let, for every $b \in \{1, \ldots, B\}$, $\Sb \in \{0, 1 \}^{ \block{\widetilde{\N}} \times \N}$ be a selection operator (\emph{i.e.}, each row only contains $0$ but a single $1$) satisfying 
    \begin{equation}\label{cond:localselection:redundance}
        \displaystyle \widetilde{\BS}_b := \underset{j \in \{1, \ldots, \block{\widetilde{\N}}\}}{\cup} \{n \in \{1, \ldots,N\} \,|\, \Sb^{(j,n)}=1\}.
    \end{equation}
    Then, $\{ \Sb \}_{1\leq \blockind\leq B}$ is called a \emph{family of local selection operators} with respect to $(\BS_b)_{1\leq \blockind\leq B}$ if
   \begin{enumerate} [label=(\roman*)]
    \item \label{cond:localselection:limited_overlap}
    for every $b\in \{1, \ldots,B\}$, $\block{\widetilde{\N}} - \Nb \mb{<} \min_{\blockind' \in \{1, \dotsc, \nblocks\}} \N_{\blockind'}$;
    
    \item \label{cond:localselection:limited_number_of_communications}
    $\max_{n\in \{1,\ldots \N\}} \card \{\blockind \in \{1, \ldots, \nblocks\} \,|\, n \in \widetilde{\BS}_b \} \mb{<} \nblocks$. 
   \end{enumerate}
\end{definition}

A few remarks can be done on Definition~\ref{prop:localselection}. 
First, it implies that $\{1, \ldots, N\} = \bigsqcup_{b=1}^B \BS_b = \bigcup_{b=1}^B \widetilde{\BS}_b$. 
Further, condition~\ref{cond:localselection:limited_number_of_communications} ensures that each element $n \in \{1, \ldots, N\}$ is only selected by a few operators $\{ \Sb \}_{1\leq \blockind\leq B}$.
Combined with \ref{cond:localselection:limited_overlap}, it further controls the size of the overlapping areas. In other words, the combination of conditions~\ref{cond:localselection:limited_overlap} and~\ref{cond:localselection:limited_number_of_communications} ensures that $(\widetilde{\BS}_b)_{1 \le b \le B}$ is close to be a partition of $\{1, \ldots, N\}$.
In a distributed computing setting, overlapping areas define the messages exchanged between the workers \mb{and in practice, the smaller $\widetilde{\N}_b - \Nb$, the better}.
Hence, to limit communication costs, we will consider \emph{localized operators} as defined below.
\begin{definition}[Localized operator] \label{prop:locality} 
    An operator $G \colon \R{\N} \to \R{\M}$ is said to be \emph{localized} if it can be decomposed into a family of local selection operators $(\Sb)_{1\leq \blockind\leq B}$ as
    \begin{equation}\label{eq:prop:locality}
        \begin{split}
            (\forall \blockind \in \{ 1, \ldots, \nblocks \})(\forall \x \in &\R{\N}) \\
            &([G(\x)]^{(i)})_{i \in \IS_b} = \block{G}(\Sb \x),
        \end{split}
    \end{equation}
    where
    $(\IS_b)_{1 \le b \le B}$ is a partition of $\{1, \ldots, M\}$ into $B$ subsets, with $M_b = \sharp\IS_b > 0$, and
    $G_b \colon \mathbb{R}^{\widetilde{\N}_b} \rightarrow \mathbb{R}^{\M_b}$.
    To simplify the notation, we use hereafter $[G(\x)]_b$ instead of $([G(\x)]^{(i)})_{i \in \IS_b}$.
\end{definition}

A localized operator $G \colon \R{N} \to \R{M}$ with respect to a family of local selection operators $(\Sb)_{1\leq \blockind\leq B}$ can be evaluated in a distributed manner (see, \emph{e.g.}, \cite{Thouvenin2024jcgs} for details). 
Let $\x \in \R{N}$ and $b \in \{1, \ldots, B\}$.
As per \eqref{eq:prop:locality}, each block $[G(\x)]_b$ only depends on the locally selected block $\Sb \x$ (or equivalently, $\IS_b$ only depends on the neighborhood $\widetilde{\BS}_b$ as given in Definition~\ref{prop:localselection}). 
In a distributed computing setting, computations underlying $[G(\x)]_b$ can be assigned to a single worker $b$, with $(x^{(n)})_{n \in \BS_n}$ directly stored in its local memory. 
To compute $[G(\x)]_b$, the worker $b$ receives $(x^{(n)})_{n \in \widetilde{\BS}_b \setminus \BS_b}$ from a small number of workers $b' \neq b$, each sending messages of small size.
The choice of the partition $(\BS_b)_{1 \leq b \leq B}$ directly impacts the local selection operators underlying the operator $G$, hence significantly affects the efficiency of the associated communications.

\begin{figure}[t]
    \centering
    \resizebox{0.9\linewidth}{!}{%
        \input{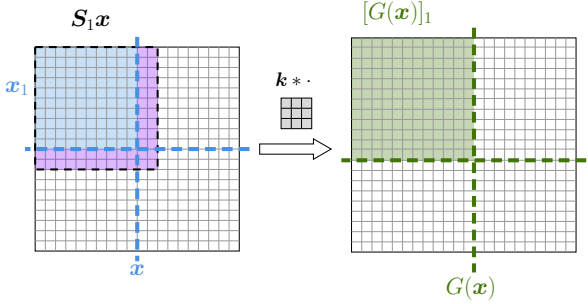}
    }

    \vspace{-0.3cm}
    
    \caption{
        Illustration of a 2D linear convolution operator $G$ defined by a $3 \times 3$ kernel $\bm{k}$, applied to an image $\bm{x}$ of size $\N = 20 \times 20$ (hence $G(\x) \in \mathbb{R}^M$ with $M = 22 \times 22$). 
        The input $\bm{x}$ is partitioned into $\nblocks = 4$ blocks (in dashed blue lines). 
        Resulting blocks in $G(\bm{x})$ are delimited by dashed green lines.
        The block $\bm{x}_1$ assigned to worker 1 is in blue, with corresponding block $[G(\bm{x})]_1$ in green.
        The purple area contains pixels received by worker $1$ to compute $[G(\bm{x})]_1$. 
        Pixels in $\bm{S}_1\bm{x}$ are in the union of the blue and purple areas.
    }
    \label{fig:locality}
\end{figure}

\begin{example}
\Cref{fig:locality} illustrates~\Cref{prop:localselection,prop:locality} with a 2D linear convolution operator defined by a kernel $\bm{k}$ of size $3 \times 3$.
An input image $\x \in \mathbb{R}^N$ of size $N = 20 \times 20$ is considered, leading to $G(\x) \in \mathbb{R}^M$ with $M = 22 \times 22$.
Dashed blue lines represent the partition of $\x$ into $\nblocks = 4$ blocks $\bm{x}_b = (x^{(n)})_{n \in \BS_b}$, with $b \in \{1, \dotsc, 4\}$.
Each block is assigned to a single worker.
The partition $[G(\bm{x})]_b = \big( [G(\bm{x})]^{(i)} \big)_{i \in \mathbb{I}_b}$ is indicated with dashed green lines.
The block $\bm{x}_1$ (in blue) is assigned to worker~$1$, as well as the corresponding output block $[G(\x)]_1$ (in green).
The pixels $(x^{(n)})_{n \in \widetilde{\BS}_1 \setminus \BS_1}$ (in purple) 
are required by worker~$1$ to compute $[G(\x)]_1$.
These pixels form a \emph{ghost region}~\cite[Section 7.5.2]{Eijkhout2023}, 
and are received from other workers before computations can be conducted on worker~$1$.
\end{example}

\smallskip\noindent\textbf{Model assumptions. ~}
Let $\nblocks \in \mathbb{N}^*$ be the number of workers available. The posterior distribution~\eqref{eq:generic_posterior_distribution} is assumed to satisfy the following assumptions, with associated notation in~\Cref{tab:notation_lin}.

\begin{assumption} \label{assumptions}\
    \begin{enumerate}[label=(A\arabic*)]
        
        \item 
        \label{assumption:lipschitz_differentiability_f1}
        The composition $f_1 \circ \Hm_1 \colon \R{\N} \to (-\infty, +\infty]$ is Lipschitz-differentiable, with Lipschitz constant $L > 0$.

        \item 
        \label{assumption:proper_lsc_f2}
        $f_2\colon \R{M_2} \to \big(-\infty, +\infty \big]$ is proper, l.s.c. and convex.

        \item 
        \label{assumption:operator_locality} 
        The linear operators $\Hm_1\colon \R{\N} \to \R{\M_1}$ and $\Hm_2\colon \R{\N} \to \R{\M_2}$ are localized in the sense of~\Cref{prop:locality}, with respect to the same partition $(\BS_{b})_{1 \le b \le B}$ of $\{1, \ldots,N\}$. \mb{Let $(\bm{S}_{i, b})_{1 \leq \blockind \leq B}$, for $i=1,2$, the corresponding families of local selection operators.}

        \item 
        \label{assumption:additive_separability}
        For $i \in \{1, 2\}$, the function $f_i \colon \R{M_i} \to \big(-\infty, +\infty \big]$ is block-additively separable
        \begin{equation} \label{eq:additive_separability}
            (\forall \z \in \R{M_i}) \quad 
            f_i(\z) = \sum_{\blockind=1}^\nblocks f_{i, \blockind} (\z_\blockind),
        \end{equation}
        where, for every $b\in \{1, \ldots, B\}$, $f_{i, \blockind} \colon \R{M_{i,\blockind}} \to \big(-\infty, +\infty \big]$, and $\z_\blockind = (z^{(m)})_{m \in \IS_{i,b}}$, with $(\IS_{i,b})_{1 \le b \le B}$ the partition of $\{1 ,\dotsc, M_i\}$ associated with $\Hm_i$ (see Table~\ref{tab:notation_lin}). 

        \item \label{assumption:locality_and_additive_separability_p} 
        There exists a family $(\bm{S}_{0, b})_{1 \leq \blockind \leq B}$ of local selection operators with respect to the partition $(\BS_b)_{1 \leq b \leq B}$ such that the function $\log p$ in \eqref{eq:generic_posterior_distribution} is block-additively separable
        \begin{equation}
            (\forall \x \in \R{N}) \quad \log p(\x) = \sum_{\blockind=1}^{\nblocks} \log p_\blockind(\bm{S}_{0, b} \x).
        \end{equation}
    \end{enumerate}
\end{assumption}

\begin{table}[!t]
    \caption{
        Notation for localized operators $\Hm_i$, with $i \in \{1, 2\}$.
    }
    \label{tab:notation_lin}
    \setlength\extrarowheight{0.1cm}
    \centering
    \begin{tabular}{c|l}
    \toprule
        $\Hm_i$ 
        &   Linear operator from $\R{\N} $ to $ \R{\M_i}$; \\
        $(\Hm_{i,b})_{1 \le b \le B}$
        &   Decomposition of $\Hm_i$ with
        $\Hm_{i, \blockind}\colon \R{\widetilde{\N}_{i,b}} \rightarrow \R{\M_{i,b}}$ and \\
        &   $[\Hm_i \x]_{\blockind} = \Hm_{i, \blockind} \bm{S}_{i,b} \x$; \\
        $   (\bm{S}_{i,b})_{1\leq \blockind\leq \nblocks}$ 
        &   Family of local selection operators; \\
        $(\BS_{b})_{1 \le b \le B}$
        &   Partition of $\{1, \ldots, N\}$ (same for both $\Hm_1$ and $\Hm_2$); \\
        $(N_{b})_{1 \le b \le B}$
        &   Cardinality for $(\BS_{b})_{1 \le b \le B}$, with $\sum_{b=1}^B N_b = N$; \\
        $(\widetilde{\BS}_{i,b})_{1 \le b \le B}$
        &   Extension of $(\BS_{b})_{1 \le b \le B}$ with overlapping; \\
        $(\widetilde{N}_{i,b})_{1 \le b \le B}$
        &   Cardinality for $(\widetilde{\BS}_{i,b})_{1 \le b \le B}$; \\
        $(\IS_{i,b})_{1 \le b \le B}$
        &   Partition of $\{1, \ldots, M_i\}$; \\
        $(M_{i,b})_{1 \le b \le B}$
        &   Cardinality for $(\IS_{i,b})_{1 \le b \le B}$, with $\sum_{b=1}^B M_{i,b} = M_i$.  \\
    \bottomrule
    \end{tabular}
\end{table}

Assumptions~\ref{assumption:operator_locality}--\ref{assumption:locality_and_additive_separability_p} are essential to the distributed sampler proposed in~\Cref{sec:methodology}, and call for a few remarks.

\begin{remark}\
\label{remark_assumption}
\begin{enumerate}
    \item 
    Assumptions~\ref{assumption:lipschitz_differentiability_f1} and \ref{assumption:proper_lsc_f2} are instrumental for Langevin-based transition kernels such as PSGLA~\cite{Salim2019}, MYULA~\cite{Durmus2018} or PnP-ULA~\cite{Laumont2022}.
    
    \item 
    Assumption~\ref{assumption:operator_locality} is motivated by the fact that if $\Hm_1$ and $\Hm_2$ are localized operators, there exists a basis where their matrix representation is block-sparse~\cite{Thouvenin2024jcgs}.
    This is satisfied for many usual inverse problems.
    Typical examples are convolutions, finite pixel differences appearing in the TV norm~\cite{Rudin1992} or masking operators used in imaging applications (see also the experiments in Section~\ref{sec:experiments}). 
    
    \item
    When the likelihood is associated with $f_i \circ \bm{H}_i$, $i = 1$ or $2$, Assumption~\ref{assumption:additive_separability} ensures that the observations $\y$ can be partitioned into $B$ conditionally-independent blocks $(y^{(m)})_{m \in \mathbb{I}_{i, b}}$ given $\x$. 
    This is in particular the case when $f_i$ comes from the likelihood function associated with an additive white Gaussian noise.
\end{enumerate}
\end{remark}

When $p$ in \eqref{eq:generic_posterior_distribution} is associated with a denoising network $D_\epsilon$, its structure has an impact not only on estimation performance, but also on speed and scalability for distributed inference.
For an efficient distributed implementation on an SPMD architecture, the structure of $D_\epsilon$ should be compatible with Assumption~\ref{assumption:locality_and_additive_separability_p} so that, after lightweight communications, it is fast to evaluate locally on each worker using subsets of locally available data.
This is the topic of the next section.


\subsection{Denoiser choice for distributed inference}
\label{sec:denoiser}

We focus on CNNs to encode the learned prior $p$.
We describe in this section how their architecture can be distributed using Assumption~\ref{assumption:locality_and_additive_separability_p}.
In our simulations, we will mainly focus on three state-of-the-art CNNs described in the examples below, namely DnCNN, DDFB and DRUNet.

\subsubsection{\textbf{Denoising CNNs}} 
\label{sec:denoiser:cnn}

Let $D_\epsilon\colon \mathbb{R}^N \to \mathbb{R}^N$ be a denoising CNN composed of $K\in\mathbb{N}^*$ layers.
In this context, $D_\epsilon$ aims to produce a good estimate of an unknown image $\overline{\x}$ from a noisy version $\bm{v} = \overline{\x} + \epsilon \w$, with $\epsilon>0$ the standard deviation of the noise, and $\w$ a realization of a standard multivariate Gaussian distribution. In other words, $D_\epsilon(\bm{v}) \approx \overline{\x}$.
In the following, the input image dimension is $N = C \times N_y \times N_x$, with $C \in \mathbb{N}^*$ the number of image channels and $N_y \times N_x$ its spatial dimensions.

The architecture of $D_\epsilon$ can generally be described as a function of $G_\epsilon = T_K \circ \cdots \circ T_1$, where for every $k\in \{1, \ldots, K\}$
\begin{equation}
    \label{eq:feedforward_cnn}
    T_k \colon \mathbb{R}^{N_k} \to \mathbb{R}^{N_{k+1}} \colon
    \bm{v}\mapsto \eta_k (\bm{W}_k\bm{v} + \bm{b}_k) .
\end{equation}
In \eqref{eq:feedforward_cnn}, the input and output dimensions are $N_k=C_k \times N_{y, k} \times N_{x, k}$ and $N_{k+1}=C_{k+1} \times N_{y, k+1} \times N_{x, k+1}$, respectively, with $N_1 = N_{K+1} = N$. 
Each layer $T_k$ contains a linear operator $\bm{W}_k \in \mathbb{R}^{M_k \times N_k}$, with $M_k = P_k \times N_{y, k} \times N_{x, k}$, 
a bias $\bm{b}_k \in \mathbb{R}^{M_k}$, 
and a non-linear activation function
$\eta_k \colon \mathbb{R}^{M_{k}} \rightarrow \mathbb{R}^{\mb{N_{k+1}}}$. 
For the sake of simplicity, in the following we directly consider the case where the linear operator $\bm{W}_k$ models a convolution. However, in practice it can also account for skip-connections, by defining $\bm{W}_k$ as the concatenation of a convolution and the identity operator (such that the current latent variable is carried forward until needed in deeper layers, \emph{e.g.}, for DRUNet described in Example~\ref{sec:denoiser:cnn:drunet}).
The convolution operator $\bm{W}_k$ can be expressed, for every $\bm{v} = (\bm{v}_c)_{1 \leq c \leq C_k} \in \mathbb{R}^{N_k}$, as 
\begin{equation}
    \label{eq:cnn:convolution}
    \bm{W}_k\bm{v} = \left( \sum_{c = 1}^{C_k} \bm{v}_c * \bm{w}_{c', c} \right)_{1 \leq c' \leq P_k } \in \mathbb{R}^{N_k},
\end{equation}
%
%
%
where $*$ is a half-padding convolution with kernel $\bm{w}_{c', c} \in \mathbb{R}^{L_y \times L_x}$ for output feature $c' \in \{1, \dotsc, P_{k} \}$ and input feature $c \in \{1, \dotsc, C_k \}$. 

\begin{example}[DnCNN denoiser \cite{Zhang2017}]\label{sec:denoiser:cnn:dncnn}
DnCNN is a simple version of \eqref{eq:feedforward_cnn}, where 
$N_2 = \cdots = N_K = P \times N_y \times N_x$, and,
for every $k \in \{1, \dotsc, \nlayers\}$, $P_k = P$. In other words, the first layer expands the input image from $C$ channels to $P$ features, keeping the same spatial dimension. All the following layers keep the same dimension, except the last layer that reduces from $P$ features to $C$ channels. It can be expressed as
\begin{equation}
    D_\epsilon(\bm{v})= (\operatorname{Id} - G_\epsilon)(\bm{v}).
\end{equation}
In general, convolution kernels are chosen of size $L_y \times L_x = 3^2$, with $K = 17$ or $20$ layers. 
The activation functions are ReLU non-linearities on all layers apart from the last, where no activation function is used (\emph{i.e.}, $\eta_{K}$ is identity).
\end{example}

\begin{example}[DDFB denoiser]\label{sec:denoiser:cnn:ddfb}
DDFB~\cite{Repetti2022eusipco} can be written as a sequential CNN with layers of the form~\eqref{eq:feedforward_cnn}, see \cite{Le2024}, and is obtained by unrolling a dual forward-backward algorithm~\cite{Combettes2010,Combettes2011}.
In a nutshell, DDFB is given by
\begin{align} 
    \label{eq:ddfb_operator}
    &D_{\epsilon}(\bm{v}) =  \proj_{[0,1]^\N} \big( \bm{v} -\gamma_{K} \bm{W}_{K}^* G_{\epsilon, \bm{v}} (\bm{W}_K \bm{v}) 
    \big),
\end{align}
with $G_{\epsilon, \bm{v}} = T_{K-1, \epsilon, \bm{v}} \circ \cdots \circ T_{1, \epsilon, \bm{v}}$.
Each layer $k \in \{1, \dotsc, K-1\}$ is the concatenation of two sub-layers: one expanding the dimension of the image from $C$ channels to $P$ features keeping the spatial dimensions fixed (\emph{via} the operator $\bm{W}_k \in \mathbb{R}^{N \times M}$, with $N = C \times N_y \times N_x$ and $M = P \times N_y \times N_x$), and one reducing back from $P$ features to $C$ channels (\emph{via} the adjoint operator $\bm{W}_k^*\in \mathbb{R}^{M \times N}$).
Specifically, for every $\bm{u} \in \mathbb{R}^{M}$,
\begin{equation}
    \label{eq:dfb_operator:T}
        T_{k, \epsilon, \bm{v}} (\bm{u}) = \mathcal{HT}_{\epsilon} \big(\uv +\gamma_k \bm{W}_k \proj_{[0,1]^\N}( \bm{v} - \bm{W}_k^* \uv) \big),
\end{equation}
where $\mathcal{HT}_{\epsilon}$ is the element-wise hard-tan operator~\cite{Le2024} with hyper-parameter $\epsilon>0$, $\proj_{\mathcal{C}}$ is a term-wise thresholding operator~\cite{Combettes2011} and $\gamma_k \in (0, 2 / \|\bm{W}_k\|^2_2)$.
In general, convolution kernels are chosen of size $L_y \times L_x = 3^2$.
\end{example}

\begin{example}[DRUNet denoiser]\label{sec:denoiser:cnn:drunet}
DRUNet~\cite{Zhang2021} is a CNN with a UNet structure which can be written as~\eqref{eq:feedforward_cnn}. 
It is composed of $K = 2I+3$ layers, including $I \in \mathbb{N}^*$ downsampling and upsampling layers. 
The layers $k \in \{2, \dotsc, K - 1\}$ include residual operators $R_k: \mathbb{R}^{N_k} \to \mathbb{R}^{N_k}$, composed of $J \in \mathbb{N}^*$ residual blocks $R_k = R_{k, J} \circ \dots \circ R_{k, 1}$.
In the following, the spatial dimensions $N_x$ and $N_y$ are assumed divisible by $2^I$ (possibly after padding), so that downsampling and upsampling operators can be applied without communications.
For $\bm{v} \in \mathbb{R}^N$, the DRUNet architecture is given by
\begin{equation}
    \label{eq:drunet}
    D_\epsilon (\bm{v}) = \bm{W}_{K} \Big( \big( G_\epsilon + \operatorname{Id} \big)\big( \bm{W}_1 [\bm{v}, \epsilon  \mathbf{1}_{N_y \times N_x}] \big) \Big),
\end{equation}
with $[\bm{v}, \epsilon  \mathbf{1}_{N_y \times N_x}] \in \mathbb R^{N_\epsilon}$ the concatenation of $\bm{v}$ and $\epsilon\mathbf{1}_{N_y \times N_x}$, $N_\epsilon = (C+1) \times N_y \times N_x$, $\bm{W}_1 \in \mathbb{R}^{M \times N_\epsilon}$ with $M = P \times N_y\times N_x$, $\bm{W}_K \in \mathbb{R}^{N \times M}$, and $G_\epsilon = T_{K-1} \circ \cdots \circ T_{2} \colon \mathbb{R}^M \to \mathbb{R}^M$.
Each layer $k \in \{2, \dotsc, I+1\}$ is defined as 
$T_{k}\colon (\bm{v}_{k'})_{2 \leq k' \leq k} \mapsto (\bm{v}_{k'})_{2 \leq k' \leq k+1}$. 
\mb{For} $k \in \{I+3, \dotsc, K-1\}$, we have $T_{k}\colon ( (\bm{v}_{k'})_{2 \leq k' \leq K-k+1}, \bm{v}_k) \mapsto ( (\bm{v}_{k'})_{2 \leq k' \leq K-k}, \bm{v}_{k+1})$ with $\bm{v}_{k'} \in \mathbb{R}^{N_{k'}}$ ($N_2=N_{K-1}=M$), and 
\begin{equation}
    \label{eq:drunet:architecture}
    \begin{split}
        & \bm{v}_{k+1} = 
        \begin{cases}
            \bm{\Phi}_k \circ R_k(\bm{v}_k), \text{ if } 2 \leq k \leq I+1, \\
            R_{I+2}(\bm{v}_{I+2}), \text{ if } k = I+2, \\
            R_k \circ \bm{\Psi}_k (\bm{v}_k + \bm{v}_{K+2-k}), \text{ if } k \geq I+3,
        \end{cases}
    \end{split}
\end{equation}
with $\bm{\Phi}_k$ and $\bm{\Psi}_k$ downsampling and upsampling operators.
The residual blocks are defined, for every $j \in \{1, \dotsc, J\}$, as
\begin{equation}
    \label{eq:drunet:residual_blocks}
    R_{k, j} = \big( \bm{W}_{k,j,2} \circ \eta_k \circ  \bm{W}_{k,j,1} + \operatorname{Id }\big)
    \colon \mathbb{R}^{N_k} \to \mathbb{R}^{N_k},
\end{equation}
where $\eta_k$ is the ReLU non-linearity, $\bm{W}_{k,j,1} \colon \mathbb{R}^{N_k} \to \mathbb{R}^{M_k}$ and $\bm{W}_{k,j,2} \colon \mathbb{R}^{M_k} \to \mathbb{R}^{N_k}$ are convolution operators with $L_y \!=\! L_x \!=\! 3$ and $P_k \!=\! C_k$.
Downsampling operators are given by
\begin{equation}
    \label{eq:drunet:downsampling}
    \bm{\Phi}_k: \bm{v} \in \mathbb{R}^{N_k} \mapsto \bm{V}_k \bm{W}_k \bm{v} \in \mathbb{R}^{N_{k+1}},
\end{equation}
where $\bm{W}_k: \mathbb{R}^{N_k} \to \mathbb{R}^{M_k}$ is a convolution operator with $L_y = L_x = 2$, $P_k = 2 C_k$, $\bm{V}_k : \mathbb{R}^{M_k} \to \mathbb{R}^{N_{k+1}}$ is a subsampling operator selecting elements at even indices along both spatial dimensions, and $N_{k+1} = (2 C_k) \times \floor{N_{y, k}/2} \times \floor{N_{x, k}/2}$.
%
Upsampling operators are given by
\begin{equation}
    \label{eq:drunet:upsampling}
    \bm{\Psi}_k: \bm{u} \in \mathbb{R}^{N_k} \mapsto \bm{W}_k \bm{U}_k \bm{u} \in \mathbb{R}^{N_{k+1}},
\end{equation}
where $L_y = L_x = 2$, $P_k = \floor{C_k / 2}$, $\bm{U}_k : \mathbb{R}^{M_k} \to \mathbb{R}^{N_{k+1}}$ inserts a zero between each consecutive entries along the spatial dimensions, and $N_{k+1} = \floor{C_k/2} \times (2N_{y, k}) \times (2N_{x, k})$.

\end{example}

\subsubsection{\textbf{Proposed distributed implementation}}
\label{sec:denoiser:distributed}

We now describe the distributed strategy proposed for~\eqref{eq:cnn:convolution}.
It relies on a Cartesian domain decomposition and communications detailed in Section~\ref{sec:denoiser:distributed:cartesian_decomposition}.
Section~\ref{sec:denoiser:distributed:forward_operator} describes the overlap-save approach~\cite{Vetterli2014} used in the proposed distributed implementation.  

\paragraph{Domain decomposition and communications}
\label{sec:denoiser:distributed:cartesian_decomposition}

\begin{figure}[t]
    \centering
    \resizebox{0.3\textheight}{!}{%
        \input{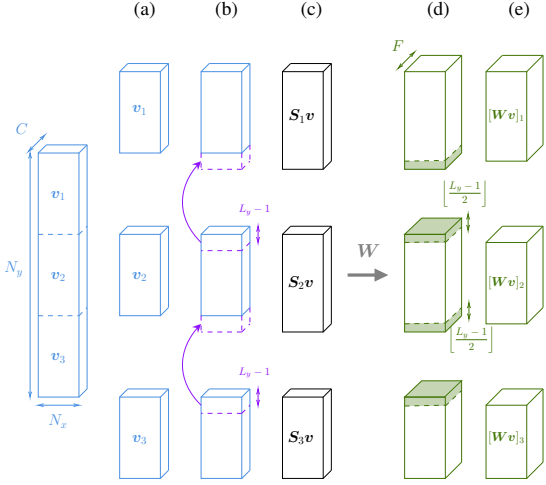}
    }
    \caption{
        Communications and operations underlying the proposed distributed implementation of $\bm{W}_k$. \mb{For readability, the layer index $k$ has been omitted from the notation.}
        Communications are represented by purple arrows, with messages in dashed purple lines.
        Discarded regions are shaded in green.
    }
    \label{fig:conv_comm_W}
\end{figure}

For the sake of simplicity, we describe a 1D domain decomposition of $\{1, \dotsc C\} \times \{1, \dotsc N_y\} \times \{1, \dotsc N_x\}$, partitioning only the axis $\{1, \dotsc N_y\}$. A 2D tessellation can be obtained using the same strategy for the axis $\{1, \dotsc N_x\}$.
To obtain a partition into blocks of similar size, we consider for Assumption~\ref{assumption:operator_locality} the sets $(\BS_b)_{1 \leq b \leq B}$ with, for every $b\in \{1, \ldots, B\}$,
\begin{multline}\label{eq:subsets_cartesian_partition}
        \BS_b 
        = \{1, \dotsc, C\}
        \times \bigg\{ \Big\lfloor \frac{(b-1) N_y}{B} \Big\rfloor + 1, \dotsc, \\
        \Big\lfloor \frac{b N_y}{B} \Big\rfloor \bigg\}
        \times \{1, \dotsc, N_x\},
\end{multline}
with $\card \BS_b = C \times N_{y, b} \times N_x$ and $N_{y, b} = \lfloor b N_y / B \rfloor - \lfloor (b-1) N_y / B\rfloor$.
The sets $(\widetilde{\BS}_b)_{1 \leq b \leq B}$ in~\eqref{eq:cnn:convolution} can be defined as
\begin{multline} \label{eq:subsets_overlapping_tessellation}
        \widetilde{\BS}_b 
        = \{1, \dotsc, C\} \times
        \bigg\{ \Big\lfloor \frac{(b-1) N_y}{B} \Big\rfloor + 1, \dotsc, \\
        \Big\lfloor \frac{b N_y}{B} \Big\rfloor + L_y - 1 \bigg\} 
        \times \{1, \dotsc, N_x\},
\end{multline}
with $\card \widetilde{\BS}_b = \widetilde{N}_{b} = C \times \widetilde{N}_{y, b} \times N_x$ and $\widetilde{N}_{y, b} = N_{y, b} + L_y - 1$.
Note that \eqref{eq:subsets_cartesian_partition}--\eqref{eq:subsets_overlapping_tessellation} satisfy \Cref{prop:localselection}~\ref{cond:localselection:limited_overlap} when $\min_{1 \leq b \leq B} N_b \geq L_y - 1$.

\Cref{fig:conv_comm_W} illustrates the partitioning and typical communication patterns in the proposed implementation.
In particular, for an image $\bm{v} \in \mathbb{R}^N$, \Cref{fig:conv_comm_W}~(a) represents the partition of $\bm{v}$ into $(\bm{v}_b)_{1 \leq b \leq B}$.
\Cref{fig:conv_comm_W}~(b) illustrates the communication pattern considered for the distributed implementation of~\eqref{eq:cnn:convolution}.
Each worker $b \in \{2, \dotsc, B\}$ is required to send a \mb{slice} covering the first $L_y - 1$ indices along axis $N_y$ (in dashed purple lines) to worker $b-1$, following the communication pattern illustrated by purple arrows.
The message received by $b \in \{1, \dotsc, B-1\}$ is stored in a ghost region~\cite[Section 7.5.2 p. 306]{Eijkhout2023} (in dashed purple lines) to form $\bm{S}_b \bm{v} = (v^{(n)})_{n \in \widetilde{\BS}_n}$.

\paragraph{Distributed implementation for $\bm{W}_k$}
\label{sec:denoiser:distributed:forward_operator}

Let $k \in \{1, \dotsc, K\}$.
The operator \eqref{eq:cnn:convolution} can be decomposed with an overlap-save approach~\cite[Section 3.9.2]{Vetterli2014} as follows.
For $\bm{v} \in \mathbb{R}^{N_k}$ and adapting notation in~\Cref{sec:denoiser:cnn} to the distributed setting from Section~\ref{sec:denoiser:distributed:cartesian_decomposition},
\begin{equation}
    \label{eq:cnn:local_decomposition}
    \bm{W}_k \bm{v} = (\bm{W}_{k,b} \bm{S}_{k,b} \bm{v})_{1 \leq b \leq B},
\end{equation}
where $\bm{S}_{k,b} \in \{0, 1\}^{\widetilde{N}_{k,b} \times N_k}$ with
$\widetilde{N}_{k,b} = C_k \times \widetilde{N}_{y, k,b} \times N_{x,k}$, $\widetilde{N}_{y, k, b} = N_{y,k,b} + L_y - 1$,
and $\bm{W}_{k,b} \in \mathbb{R}^{M_{k,b} \times \widetilde{N}_{k,b}}$ 
with $M_{k,b} = P_k \times N_{y, k, b} \times N_{x,k}$.
More precisely, for $\widetilde{\bm{v}} = (\widetilde{\bm{v}}_{c})_{1 \leq c \leq C_k} \in \mathbb{R}^{\widetilde{N}_{k,b}}$,
\begin{equation}
    \label{eq:cnn:local_convolution}
    \bm{W}_{k,b} \widetilde{\bm{v}} = \left( \sum_{c = 1}^{C_k} \bm{A}_{k, b} (\widetilde{\bm{v}}_{c} * \bm{w}_{c', c}) \right)_{1 \leq c' \leq P_k },
\end{equation}
with $\bm{A}_{k,b} \in \{0, 1\}^{(N_{y,k,b} \times N_{x,k}) \times (\widetilde{N}_{y, k, b} \times N_{x,k})}$ a local selection operator.
\Cref{fig:conv_comm_W}~(b)--(d) illustrates the operators in~\eqref{eq:cnn:local_decomposition}--\eqref{eq:cnn:local_convolution}.
After communications (see \Cref{fig:conv_comm_W}~(b)), each worker $b$ can access $\bm{S}_{k,b} \bm{v}$ (\Cref{fig:conv_comm_W}~(c)) for~\eqref{eq:cnn:local_convolution}.
Entries computed with the wrong boundary conditions are discarded via $\bm{A}_{k,b}$ (\Cref{fig:conv_comm_W}~(d)), leading to the output blocks in \Cref{fig:conv_comm_W}~(e).

In the following, we propose to distribute the computations of each successive layer $k \in \{1, \dotsc, \nlayers\}$, communicating the required boundary elements to compute local convolutions~\eqref{eq:cnn:local_convolution}. 
In practice, the direction of communications in~\Cref{fig:conv_comm_W}~(b) is reversed from one convolution operator to another to maintain balanced workloads and memory usage for all workers over the sequence of convolutional layers.

Other distributed strategies could be contemplated, such as communicating the receptive field of the full sequence of convolutions computed across the network~\cite{Galerne2024}.
This alleviates the need for successive communications phases in each layer, at the cost of a significant increase in the size of messages as the input dimensions and the number of layers increase.
Comparing different communication patterns is an interesting perspective beyond the scope of this paper.


\subsection{Proposed distributed sampler}
\label{sec:dsgs_pnp_ula_psgla}

Following the AXDA approach~\cite{Vono2020}, we approximate~\eqref{eq:generic_posterior_distribution} by introducing an auxilliary variable $\z \in \R{\M_2}$ as
\begin{multline} \label{eq:general_posterior_l2_axda}
        \pi_\rho(\x, \z | \y) 
        \propto p(\x) \exp \Big(- f_1(\Hm_1\x) - f_2(\z) 
        \\
        - \frac{1}{2\rho}\|\z - \Hm_2\x\|^2_2 \Big),
\end{multline}
where $\rho > 0$ controls the coupling between $\z$ and $\Hm_2\x$. Note that the same technique could be applied to $f_1$ as well.
Samples can be drawn from~\eqref{eq:general_posterior_l2_axda} 
with SGS~\cite{Vono2019}, where the associated conditional distributions are
\begin{equation} \label{eq:sgs:marginal_x}
    \pi_{\rho}(\x | \y, \z) \propto p(\x) \exp\big( 
        - f_1(\Hm_1\x) 
        - \frac{1}{2\rho}\|\z - \Hm_2\x\|^2_2 
    \big),
\end{equation}
\begin{equation} \label{eq:sgs:marginal_z}
  \pi_{\rho}(\z | \y, \x) \propto \exp\big(-f_2(\z) - \frac{1}{2\rho}\|\z - \Hm_2\x\|^2_2 \big).
\end{equation}
In practice, \eqref{eq:sgs:marginal_x} and \eqref{eq:sgs:marginal_z} can be significantly less expensive to sample from compared to~\eqref{eq:general_posterior_l2_axda}.
Under 
\mb{Assumptions~\ref{assumption:lipschitz_differentiability_f1} and~\ref{assumption:proper_lsc_f2}}, PSGLA~\cite{Salim2019} can be used to sample~\eqref{eq:sgs:marginal_z}.
Further, to leverage a pre-trained prior, we propose to use the \mb{transition kernel underlying the} PnP Unadjusted Langevin Algorithm (PnP-ULA)~\cite{Laumont2022} to update $\x^{(t+1)}$ in~\eqref{eq:sgs:marginal_x}. 
\mb{This approach} has the advantage that it ensures the existence of a proper prior distribution with pdf $p$ in~\eqref{eq:generic_posterior_distribution} associated with a denoiser $D_{\epsilon}$ (pre-trained at noise level $\epsilon > 0$), assuming $\bm{I}_{\N} - D_{\epsilon}$ is Lipschitz continuous with constant $L_\epsilon > 0$~\cite[Section 3.2]{Laumont2022}.

Hence, for $\kappa \in (0, \rho)$, the transition from iteration $t \in \mathbb{N}^*$ to $t+1$ can be written
\begin{align} 
    \x^{(t+1)} 
    &   = \x^{(t)}
        - \gamma \Hm_1^* \nabla f_1 (\Hm_1\x^{(t)})  \nonumber\\
        & \quad 
        - \frac{\gamma}{\rho} \Hm_2^* (\Hm_2\x^{(t)} - \upd{\z}{t}) 
        + \frac{\alpha\gamma}{\epsilon^2} \big( D_\epsilon (\x^{(t)}) - \x^{(t)} \big) \nonumber \\
        & \quad 
        + \frac{\gamma}{\lambda} \big( \proj_{\mathcal{C}}(\x^{(t)}) - \x^{(t)} \big) + \sqrt{2\gamma} \bm{\xi}^{(t+1)}, 
    \label{eq:sgs_pnp_ula_psgla:pnp_ula}\\
    \upd{\z}{t+1} 
    &   = \prox_{\kappa f_2} \big( \upd{\z}{t} 
        - \frac{\kappa}{\rho} \big(\upd{\z}{t} - \Hm_2\x^{(t+1)}\big) \nonumber\\[-0.2cm]
        & \quad \quad \quad \quad \quad \quad \quad \quad \quad 
        + \sqrt{2\kappa} \bm{\zeta}^{(t+1)} \big),    \label{eq:sgs_pnp_ula_psgla:psgla}
\end{align}
where $\bm{\xi}^{(t+1)} \sim \mathcal{N}(\mathbf{0}, \mathbf{I}_\N)$, $\bm{\zeta}^{(t+1)} \sim \mathcal{N}(\mathbf{0}, \mathbf{I}_{M_2})$,
and $\lambda, \alpha, \gamma > 0$ are such that 
\begin{equation}\label{eq:stepsize_cond}
    \begin{cases}
         2(L + \|\Hm_2\|_2^2) + \alpha\epsilon^{-2} L_\epsilon  \leq (2\lambda)^{-1}, \\
        3\gamma (L + \|\Hm_2\|_2^2 + \lambda^{-1} + \alpha \epsilon^{-2}L_\epsilon ) < 1 .
    \end{cases}
\end{equation}

\begin{algorithm}[tbp]
    \caption{Proposed distributed sampler.}
    \label{algo:dsgs_pnp_ula_psgla}
    \footnotesize
    \KwIn{%
        $\y \in \R{\M}$, $\epsilon, \rho > 0$, and $\kappa \in (0, \rho)$. 
        Choose 
        $\alpha, \lambda, \gamma > 0$ satisfying \eqref{eq:stepsize_cond}. 
    } 
    \For(in parallel){each worker $\blockind \in \{1, \dotsc, \nblocks \}$}{
        Load and store observation block $\yb \in \R{\M_{1, \blockind}}$\;
        Initialize $\xb^{(0)} \in \R{\Nb}$ and $\zb^{(0)} \in \R{\M_{1, \blockind}}$\;

        \For{$t = 0$ \KwTo $T - 1$}{
            \tcp{Update $\xb$ with PnP-ULA \mb{transition}~\cite{Laumont2022} (see \eqref{eq:sgs_pnp_ula_psgla:pnp_ula})}

            Communicate to retrieve $\big( \bm{S}_{i, b}\x^{(t)} \big)_{0 \leq i \leq 2}$\;
            \label{line:grad_x:comm_forward}

            $\vect{u}_{1,b}^{(t)} = \Hm_{1, b}^* \nabla f_{1, b} (\Hm_{1, b}\bm{S}_{1, b}\x^{(t)})$\;
            \label{line:grad_x:local_computations_H1}
            
            $\vect{u}_{2,b}^{(t)} = \frac{1}{\rho} \Hm_{2, b}^* (\Hm_{2, b}\bm{S}_{2, b}\x^{(t)} - \zb^{(t)})$\;
            \label{line:grad_x:local_computations_H2}
            \smallskip

            Communicate to compute $\displaystyle \vect{v}_b^{(t)} = \sum_{j=1}^B \bm{S}^*_{1,j} \vect{u}_{1,b}^{(t)} + \bm{S}^*_{2,j} \vect{u}_{2,j}^{(t)}$\;
            \label{line:grad_x:comm_backward}
            \smallskip
            
            Communicate \& compute $\blockp{D_{\epsilon}(\upd{\x}{t})}$ (see \Cref{sec:denoiser})\; \label{line:ddfb}
            \smallskip

            Sample $\vect{\xi}_\blockind^{(t+1)} \sim \mathcal{N}(\mathbf{0}, \mathbf{I}_{\N_\blockind})$\;
            \label{line:pnp_ula:gaussian}

            $\x_\blockind^{(t+1)} = \xb^{(t)} - \gamma \vect{v}_\blockind^{(t)} 
            + \frac{\alpha \gamma}{\epsilon^2} \big( \blockp{D_{\epsilon}(\upd{\x}{t})} - \xb^{(t)} \big)
            + \frac{\gamma}{\lambda} \big( \proj_{\mathcal{C}_\blockind}(\xb^{(t)}) - \xb^{(t)} \big)
            + \sqrt{2\gamma} \vect{\xi}_\blockind^{(t+1)}$\;
            \label{line:upd_x}
            \medskip

            \tcp{Update $\zb$ with PSGLA \mb{transition}~\cite{Salim2019} (see \eqref{eq:sgs_pnp_ula_psgla:psgla})}

            Communicate to retrieve $\bm{S}_{2, b}\bm{x}^{(t+1)}$\; 
            \label{line:comm_H2}

            Sample $\vect{\zeta}_\blockind^{(t+1)} \sim \mathcal{N}(\mathbf{0}, \mathbf{I}_{M_{2,b}})$\;
            \label{line:psgla:gaussian}

            $\zb^{(t+1)} = \prox_{\kappa f_{2,\blockind}} \big( \vect{\z}_\blockind^{(t)} - \frac{\kappa}{\rho} \big( \z_\blockind^{(t)} - \bm{H}_{2,b} \bm{S}_{2, b}\bm{x}^{(t+1)} \big)$ 
            $\qquad + \sqrt{2\kappa} \vect{\zeta}_\blockind^{(t+1)} \big)$\; \label{line:upd_z} 
        }
    }
    \KwOut{($\upd{\x}{t}, \upd{\z}{t})_{1 \leq t \leq T}$}
\end{algorithm}

Note that variants of the proposed sampler~\eqref{eq:sgs_pnp_ula_psgla:pnp_ula}--\eqref{eq:sgs_pnp_ula_psgla:psgla} can be obtained with other combinations of a PnP-based transition such as~\cite{Faye2025}, with a Langevin kernel relying on a proximal operator, such as MYULA~\cite{Durmus2018}, PSGLA~\cite{Salim2019} or SK-ROCK~\cite{Pereyra2020}.

The distributed setting of~\Cref{sec:prob_statement} paves the way to a distributed counterpart to~\eqref{eq:sgs_pnp_ula_psgla:pnp_ula}--\eqref{eq:sgs_pnp_ula_psgla:psgla} inspired by~\cite{Thouvenin2024jcgs}, where the denoiser $D_\epsilon$ is also distributed, as described in \Cref{sec:denoiser}.
The resulting distributed sampler is reported in~\Cref{algo:dsgs_pnp_ula_psgla}.

Under~\ref{assumption:operator_locality}--\ref{assumption:additive_separability}, computations associated with $f_1$, $f_2$, $\Hm_1$ and $\Hm_2$ can be decomposed over $\nblocks \geq 2$ workers. 
In addition, according to \cite[Section~3.2]{Laumont2022}, there exists a proper prior distribution with pdf $p$ in~\eqref{eq:generic_posterior_distribution} associated with the denoiser $D_{\epsilon}$.
Further, following the proposed distributed strategy for $D_\epsilon$ described in \Cref{sec:denoiser:distributed}, the associated prior $p$ satisfies Assumption~\ref{assumption:locality_and_additive_separability_p}.
This allows the PnP-ULA (\Cref{algo:dsgs_pnp_ula_psgla} lines \ref{line:grad_x:comm_forward}--\ref{line:upd_x}) and PSGLA (lines \ref{line:comm_H2}--\ref{line:upd_z}) steps to be distributed on an SPMD architecture.

\begin{remark}
    A few remarks can be made on the distributed strategy considered in \Cref{algo:dsgs_pnp_ula_psgla}.
    \begin{enumerate}
        \item 
        Most steps in \Cref{algo:dsgs_pnp_ula_psgla} do not require synchronization.
        \item 
        Communications are only required between a few workers in lines~\ref{line:grad_x:comm_forward}, \ref{line:grad_x:comm_backward}, \ref{line:ddfb} and \ref{line:comm_H2}.
        The distributed PnP-ULA transition induces communications to compute $\nabla (f_1 \circ \bm{H}_1)$ (lines~\ref{line:grad_x:comm_forward} and \ref{line:grad_x:comm_backward}) and apply the denoiser (line~\ref{line:ddfb}).
        Some entries in $\x^{(t+1)}$ are exchanged between a few workers in line~\ref{line:comm_H2} so that $\z$ can be updated locally (line~\ref{line:upd_z}).
        \item 
        The messages in line~\ref{line:grad_x:comm_forward} can be grouped into a single message to reduce communication costs.
    \end{enumerate}
\end{remark}


\section{Image restoration experimental setting} 
\label{sec:experiments}

The proposed~\Cref{algo:dsgs_pnp_ula_psgla} is assessed on three imaging tasks detailed in~\Cref{sec:applications}, using three sets of experiments described in~\Cref{sec:exp_settings}.
Codes to reproduce the experiments will be made available at \url{https://github.com/maxime-bouton/cards}. 
All the experiments have been conducted on the CNRS supercomputer Jean Zay, hosted by GENCI at IDRIS. Algorithms were run on computing nodes equipped with two 2.5 GHz, 20-core Intel Cascade Lake 6248 processors, and four Nvidia Tesla V100 SXM2 GPUs with 32 GB of memory each.

\subsection{Application setting} \label{sec:applications}

Three different applications are considered: image inpainting under additive white Gaussian noise, and image deconvolution under either additive white Gaussian noise or Poisson noise.
Synthetic observations have been generated from high-resolution images taken from \url{https://pixabay.com}.
Ground truth images $\widebar{\bm{x}}$ have been cropped to pre-defined sizes $\N$ (see~\Cref{sec:exp_settings}), and normalized into $\mathcal{C} = [0, 1]^N$.
Estimates and pixel-wise variance have been computed from $T = 10^4$ samples from~\Cref{algo:dsgs_pnp_ula_psgla}, including $10^3$ burn-in samples.

\subsubsection{\textbf{Inpainting with Gaussian noise}}
\label{sec:gaussian_inpainting}

The likelihood is
\begin{equation} \label{eq:likelihood:gaussian_inpainting}
    \y \mid \x \sim \mathcal{N}(\bm{H}_1 \x, \sigma^2 \bm{I}_{N}), 
\end{equation}
with $\bm{H}_1 = \bm{H} \in \{0,1\}^{N \times N}$ a diagonal operator encoding the selection mask, and $\sigma^2 > 0$ the noise variance.
Observations $\bm{y}$ are generated with $M=0.3 N$ observed pixels and a noise variance $\sigma^2$ leading to a 15~dB input SNR.

The pdf of the posterior distribution is of the form~\eqref{eq:generic_posterior_distribution}, with
\begin{equation}
\label{eq:potential_gaussian_likelihood}
    f_1(\bm{H}_1 \bm{x}) = \frac{1}{2\sigma^2} \| \y - \bm{H}\bm{x} \|_2^2,
\end{equation}
where the prior is handled through $p$ or $f_2\circ \bm{H}_2$.
The masking operator acts element-wise, so no communication is needed to evaluate $\Hm_1$ and $\Hm_1^*$ in a distributed computing setting, \emph{i.e.}, $M_{1,\blockind} = N_\blockind$ and $\bm{S}_{1, \blockind} = \bm{I}_{N_\blockind}$ in~\eqref{eq:additive_separability}.

\subsubsection{\textbf{Deconvolution with Gaussian noise}}
\label{sec:gaussian_deconvolution}

The likelihood is also of the form~\eqref{eq:likelihood:gaussian_inpainting} with $\bm{H}_1 = \bm{H} \in \mathbb{R}^{M \times N}$ a linear convolution operator.
A normalized motion blur is considered for $\bm{H}$, with kernel size $L_x \times L_y$ increasing in the same proportion as $\N$: $(\N, L_xL_y) \in \{ (3\times2048^2, 65^2), (3\times2896^2, 91^2), (3\times4096^2, 129^2)\}$.
Observations are generated with a noise variance $\sigma^2$ such that the input SNR is 25~dB.

The pdf of the posterior distribution is of the form~\eqref{eq:generic_posterior_distribution}, \eqref{eq:potential_gaussian_likelihood}, with the prior handled through $p$ or $f_2\circ \bm{H}_2$.
Communications similar to~\Cref{sec:denoiser} are required to compute $\bm{H}_1$ and $\bm{H}_1^*$ in a distributed computing setting.

\subsubsection{\textbf{Deconvolution with Poisson noise}}
\label{sec:poisson_deconvolution}

The likelihood is
\begin{equation} \label{eq:likelihood:poisson_deconvolution}
    \y \mid \x \sim \mathcal{P}(\eta \bm{H} \x),
\end{equation}
with $\mathcal{P}$ the Poisson distribution, $\bm{H}_1 = \bm{H} \in \mathbb{R}^{M \times N}$ a linear convolution operator, and $\eta > 1$ controlling the variance of the Poisson noise.
Observations have been generated with $\eta = 250$ and the same convolution kernel as in~\Cref{sec:gaussian_deconvolution}.

The pdf of the posterior distribution is of the form~\eqref{eq:generic_posterior_distribution}, where $f_1 \circ \bm{H}_1 \equiv 0$, and $f_2 \circ \bm{H}_2$ is split to handle both the pdf of the likelihood and part of the prior (see \Cref{ssec:setting:prior}), \emph{i.e.},
\begin{equation}\label{eq:poisson:f2}
    f_2(\bm{H}_2 \bm{x}) = \KL(\y \| \eta \bm{H}\x) + \widetilde{f}_2(\widetilde{\bm{H}}_2 \x),
\end{equation}
with $\KL$ being the Kullback-Leibler divergence. 
Both $\widetilde{f}_2 \circ \widetilde{\bm{H}}_2$ and $p$ will be specified in \Cref{ssec:setting:prior}.

As in~\cite{Thouvenin2024jcgs}, two blocks are used in the AXDA auxiliary variable $\bm{z} = (\z_1^T, \z_2^T)^T$, with coupling parameters $\rho_1, \rho_2 > 0$.

\subsection{Choice of the prior and algorithm setting}
\label{ssec:setting:prior}

Experiments are conducted for the three applications described in~\cref{sec:applications}, using two types of priors, both satisfying~\ref{assumption:additive_separability}--\ref{assumption:locality_and_additive_separability_p}.
In our simulations, the sampler is initialized as follows. Auxiliary variables $\bm{z}$ are initialized to $\bm{0}$. For the inpainting problem in~\Cref{sec:gaussian_inpainting}, $\bm{x}$ is initialized by interpolating $\bm{y}$ with bicubic splines. For the deconvolution problems in~\Cref{sec:gaussian_deconvolution,sec:poisson_deconvolution}, $\bm{x}$ is initialized to $\bm{0}$.

\subsubsection{\textbf{Denoiser-based prior}}
\label{item:prior_choice:denoisers}
    
We consider three denoising CNNs, namely DnCNN, DDFB and DRUNet, described in Examples~\ref{sec:denoiser:cnn:dncnn}, \ref{sec:denoiser:cnn:ddfb} and \ref{sec:denoiser:cnn:drunet}, respectively.

Both DnCNN and DRUNet are pre-trained denoisers, whose weights and implementation are available at \url{https://github.com/cszn/KAIR}.
The DnCNN network contains $\nlayers = 20$ layers, each with $F = 64$ filters.
The DRUNet network has been trained with $I = 3$ downsampling/upsampling layers, $J=4$ blocks per residual layer, and $F = 64$ features.
The DDFB network contains $\nlayers = 4$ layers, with $F = 64$ latent features.
We trained the networks using an $\ell_1$-loss, on random patches of size $50 \times 50$ from the ImageNet test set~\cite{Russakovsky2015}, corrupted by additive white Gaussian noise with standard deviation $\epsilon$ selected uniformly at random between $0$ and $0.1$. 
Training has been carried out over 100 epochs using the Adam optimizer~\cite{Kingma2017}, with a learning rate of $10^{-3}$, a weight decay of $10^{-4}$ and a batch size of $10^{3}$.
The Lipschitz constant $L_\epsilon$ of $\bm{I}_N - D_\epsilon$ has been estimated for each denoiser as in~\cite{Pesquet2021} (see Table~\ref{tab:denoising_comparison_runtime} for values).

Depending on the noise model (see \Cref{sec:applications}), the functions $p$ and $f_2 \circ \bm{H}_2$ can be identified as follows.
\begin{enumerate}[label=(\roman*)]
    \item \textbf{Gaussian noise}: 
    We define $f_2 \circ \bm{H}_2 \equiv 0$, and $p$ as in~\cite{Laumont2022}.
    The posterior distribution can be directly sampled with PnP-ULA~\cite{Laumont2022} without using AXDA.
    In the experiments, $\alpha=1$, $\epsilon=\sigma$, $\lambda = 0.99/(4 \|\bm{H}\|_2^2/\sigma^2 + 2\alpha\epsilon^{-2}L_\epsilon)$ and $\gamma = 0.99/(\alpha L_\epsilon/\epsilon^2 + \|\eta \bm{H}\|_2^2/\rho + \lambda^{-1})/3$.

    \item \textbf{Poisson noise}: 
    As the likelihood~\eqref{eq:poisson:f2} involves a KL divergence, a hybrid regularization is adopted to enforce non-negativity on auxiliary variables after applying AXDA.
    We define $p$ as in~\cite{Laumont2022}, $\widetilde{\bm{H}}_2  = \bm{I}_{N}$ and $\widetilde{f}_2 = \iota_{\mathbb{R}_+^N}$.
    AXDA is applied as in~\eqref{eq:general_posterior_l2_axda}, and the posterior distribution is sampled with~\Cref{algo:dsgs_pnp_ula_psgla}.
    \mb{The PSGLA transition kernel} is used to update $\bm{z} = (\z_1^T, \z_2^T)^T$ with stepsizes $\kappa_1, \kappa_2 > 0$, respectively.
    In the experiments, $\alpha=1$, $\epsilon=0.05$, $\rho_1 = 10$, $\rho_2=10^{-3}$, $\lambda = 0.99/(4 \|\eta\bm{H}\|_2^2/\rho_1 + 4\rho_2^{-1} + 2\alpha \epsilon^{-2}L_\epsilon)$, $\gamma = 0.99/(\alpha L_\epsilon/\epsilon^2 + \|\eta \bm{H}\|_2^2/\rho_1 + \rho_2^{-1} + \lambda^{-1})/3$, $\kappa_1 = 0.99\rho_1$ and $\kappa_2 = 0.99\rho_2$.    
\end{enumerate}

\subsubsection{\textbf{Isotropic total variation (TV) prior~\cite{Rudin1992}}}
\label{item:prior_choice:tv}
    
In this case the pdf of the prior is of the form of
\begin{equation*}
    \label{eq:tv_prior}
    (\forall \x \in \mathbb{R}^N) \quad 
    \pi(\x) \propto \exp\big( -\beta \|\bm{D}\x\|_{2, 1} \big),
\end{equation*}
with $\beta > 0$ a regularization parameter, $\bm{D}: \mathbb{R}^N \rightarrow \mathbb{R}^{N \times 2}$ the 2D discrete gradient operator, and
\begin{equation*}
    (\forall \z = (\z_n)_{1 \leq n \leq N} \in \mathbb{R}^{N \times 2}) \quad 
    \|\z\|_{2, 1} = \sum_{n = 1}^N \|\z_n\|_2.
\end{equation*}

The functions $p$ and $f_2 \circ \bm{H}_2$ can be identified as follows.
\begin{enumerate}[label=(\roman*)]
    \item \textbf{Gaussian noise}: 
    In this case, $p = \mathds{1}_{\mathbb{R}_+^N}$, 
    $f_2 = \beta \|\cdot\|_{2, 1}$ and $\bm{H}_2 = \bm{D}$.
    The AXDA approach is applied as in~\eqref{eq:general_posterior_l2_axda}.
    The variable $\bm{x}$ is updated with \mb{the PSGLA~\cite{Durmus2018} transition kernel} with stepsize $\gamma > 0$. 
    In the experiments, $\rho=10^{-5}$, $\beta=40$, $\gamma = 0.99 (\|\bm{H}_1\|_2^2 / \sigma^2 + \|\bm{D}\|_2^2 /\rho )^{-1}$, $\kappa = 0.99 \rho \|\bm{D}\|_2^{-2}$.
    
    \item \textbf{Poisson noise}: 
    We consider $p = \mathds{1}_{\mathbb{R}_+^N}$, and $f_2\circ \bm{H}_2$ as in \eqref{eq:poisson:f2} with $\widetilde{\bm{H}}_2  = \bm{D}$ and $\widetilde{f}_2 =  \beta \| \cdot\|_{2, 1}$.
    AXDA is applied as in~\eqref{eq:general_posterior_l2_axda}.
    The variables $\bm{x}$ and $\bm{z} = (\bm{z}_1^T, \bm{z}_2^T)^T$ are updated with \mb{the PSGLA transition kernel}~(see~\cite{Thouvenin2024jcgs}), with stepsizes $\gamma$, $\kappa_1$ and $\kappa_2 > 0$, respectively.
    In the experiments, $\beta=13$, $\rho_1 = 10$, $\rho_2=10^{-3}$, $\gamma = 0.99/(\|\eta \bm{H}\|_2^2 / \rho_1 + \|\bm{D}\|_2^2/\rho_2)$, $\kappa_1 = 0.99\rho_1$ and $\kappa_2 = 0.99 \rho_2$.
\end{enumerate}

\subsection{Evaluation setting} \label{sec:exp_settings}

Performance is evaluated with the following experiments.
\begin{enumerate}
    \item \textbf{Performance and communication cost analysis}: 
    Denoising performance, computation and communication costs of the denoisers are reported in~\cref{sec:exp_results:costs_denoisers}. Denoising performance is evaluated on $1000$ randomly selected $50\times50$ patches from DIV2K~\cite{agustssonNTIRE2017Challenge2017}, corrupted with a Gaussian noise so that the input SNR is $20$ dB.

    \item \textbf{Restoration quality experiments}: Estimation quality 
    is discussed in~\Cref{sec:exp_results:quality}, for $\nblocks \in \{1, 2, 4\}$ GPUs and the different choices of priors considered.
    %
    Reconstruction quality is evaluated with the minimum mean square error (MMSE) estimator.
    Quality of the estimates is quantified in terms of structural similarity index (SSIM)~\cite{Wang2004ssim}, peak signal-to-noise ratio (PSNR)~\cite{wangMeanSquaredError2009}, and SNR defined by
    \begin{equation}
        \label{eq:rsnr}
        \snr(\overline{\x}, \widehat{\x}) = 10 \log_{10} \Big( \frac{\|\overline{\x}\|_2^2}{\|\overline{\x} - \widehat{\x}\|_2^2} \Big),
    \end{equation}
    with $\overline{\x}$ the ground truth image and $\widehat{\x}$ an estimate.
    Due to memory constraints, uncertainty quantification is reported using the pixel-wise variance of the chain after burn-in\footnote{Pixel-wise variance can be computed while generating the chain, whereas 95\% credible intervals require post-processing the samples stored to disk.}.

    \item \textbf{Scalability experiments}: Strong and weak scaling performance of~\Cref{algo:dsgs_pnp_ula_psgla} are evaluated with one CPU core allocated to each of the $\nblocks$ GPUs leveraged in the algorithm. 
    Both are assessed in terms of average runtime per iteration.
    Strong scaling is assessed with $\nblocks \in \{1, 2, 4\}$ GPUs to restore an image of size $\N = 3 \times 2048 \times 2048$.
    Weak scaling is evaluated on problems with image sizes and resources increasing in the same ratio, using $C = 3$ and $(N_yN_x, \nblocks) \in \{ (2048^2, 1), (2896^2, 2), (4096^2, 4)\}$.
    Results are reported and discussed in \Cref{sec:exp_results:scalability}.
\end{enumerate}



\begin{table}[tbp]
    \caption{
        \mb{\textbf{Denoiser comparison -- }}
        Number of parameters, memory footprint, \mb{estimation of the} Lipschitz constant $L_\epsilon$ for $(\bm{I}_{n} - D_\epsilon)$ and \mb{serial} runtime.
    }
    \label{tab:denoising_comparison_runtime}

    \centering
    \footnotesize
    \begin{tabular}{lrrrc}
        \toprule
        \textbf{Denoiser} & \textbf{\#Params} & \textbf{Mem. (MiB)} & $L_\epsilon$ & \textbf{Time (ms)}\\ \midrule
        \textbf{DDFB ($K=4$)} & 6\,912 & 0.03 & 2 & 56.44\\
        \textbf{DDFB ($K=20$)} & 34\,560 & 0.13 & 2 & 312.48\\
        \textbf{DnCNN} & 668\,227 & 2.55 & 3 & 406.27\\
        \textbf{DRUNet} & 32\,640\,960 & 124.52 & 7 & 885.86\\
        \bottomrule
    \end{tabular}
\end{table}

\begin{table}[tb]
    \caption{\mb{\textbf{Gaussian denoising performance -- }}
    Comparison between denoisers in terms of output quality ($B = 1$ GPU).}
    \label{tab:denoising_comparison_metrics}

    \centering
    \footnotesize
    
    \begin{tabular}{lccc}
        \toprule
        \textbf{Denoiser} & \textbf{SNR} & \textbf{PSNR} & \textbf{SSIM} \\ \midrule
        \textbf{Noisy input} & $20.00 \pm 0.07$ & $27.80 \pm 3.95$ & $0.689 \pm 0.196$ \\ \midrule
        \textbf{DDFB ($K=4$)} & $27.24 \pm 3.09$ & $35.04 \pm 4.18$ & $0.928 \pm 0.041$ \\
        \textbf{DDFB ($K=20$)} & $27.61 \pm 3.27$ & $35.41 \pm 4.24$ & $0.935 \pm 0.038$ \\
        \textbf{DnCNN} & $28.54 \pm 3.40$ & $36.10 \pm 3.90$ & $0.948 \pm 0.031$ \\
        \textbf{DRUNet} & $29.13 \pm 3.74$ & $36.69 \pm 4.19$ & $0.953 \pm 0.028$ \\ \bottomrule
    \end{tabular}
\end{table}

\section{Experimental results} \label{sec:exp_results}

\subsection{Performance and communication cost analysis} \label{sec:exp_results:costs_denoisers}

\begin{table}[t]
    \caption{
        \mb{\textbf{Communication evaluation --}}
        Number of floating point operations (FLOPs), 
        message sizes and communications per worker to distribute the different priors when partitioning only the axis $\{1, \dotsc, N_y\}$ with $\nblocks \geq 2$ workers.
    }
    \label{tab:denoiser:costs}
    \centering
    \resizebox{0.48\textwidth}{!}{%
    \begin{tabular}{lrrr}\toprule
    \textbf{Prior} & \textbf{FLOPs} & \textbf{Message size} & \textbf{$\sharp$ Comms.} \\ 
    \midrule
    \textbf{TV} & $6 N_x(N_y/B + 1)$ & $6N_x$ & $1$\\ 
    \midrule
    \textbf{DDFB} ($K=4$) & $27\,648 N_x(N_y/B + 2)$ & $ 67\N_x$ & $8$\\ 
    \midrule
    \textbf{DnCNN} & $1\,336\,451 N_x(N_y/B + 2)$ & $122\N_x$ & $20$\\ 
    \midrule
    \textbf{DRUNet} & $4\,236\,054 N_x(N_y/B + 2)$ & $126\N_x$ & $58$\\ 
    \bottomrule
    \end{tabular}
    }
\end{table}

\Cref{tab:denoising_comparison_runtime} compares the denoisers from~\Cref{sec:denoiser} in terms of number of parameters, memory footprint, Lipschitz constant $L_\epsilon$ and runtime with $B = 1$ worker.
As expected, heavier networks such as DnCNN and DRUNet contain more parameters, have a larger Lipschitz constant and require more runtime.
\Cref{tab:denoising_comparison_metrics} reports the associated performances for Gaussian denoising.
It highlights the trade-off between network complexity and performance.
In particular, the limited improvements in quality obtained with additional DDFB layers may not justify the corresponding increase in runtime and memory. 

\Cref{tab:denoiser:costs} compares the computation and communication costs of the different denoisers using the proposed distributed implementation (see~\Cref{sec:denoiser:distributed}).
Computation load is reported as the maximum amount of floating-point operations (FLOPs) carried out by a worker, estimated numerically with the \texttt{calflops} library\footnote{Available at \url{https://github.com/MrYxJ/calculate-flops.pytorch}}. 
Message sizes are reported as the average number of elements a worker needs to send in the distributed setting from~\Cref{sec:denoiser:distributed}, which is independent from the number of workers $\nblocks$.
Note that splitting along both spatial dimensions $N_x$ and $N_y$ would lead to comparable computing loads and message sizes, with differences mainly in the number of communications (with up to 4 neighbour workers, instead of 2 when partitioning along a single axis).

Although DDFB requires far fewer computations and parameters than the other NNs, its communication costs remain comparable.
In fact, only narrow border regions need to be exchanged between workers (1 pixel wide for TV, against $L_y - 1 = 2$ for the CNNs).
Dominant factors in communication costs are the number of channels of latent variables for message sizes, and the number of convolutional layers for the number of communication phases.

For DRUNet, although the number of channels doubles after each downsampling layer, both spatial dimensions $N_x$ and $N_y$ are divided by 2: the resulting computation load is thus divided by 2. 
For $N_x$, $N_y$ and $B$ fixed, the ratio of FLOPs to message size per worker is similar for DnCNN and DRUNet, and approximately 10 times larger than DDFB.
This suggests that DnCNN and DRUNet are more computationally-intensive relative to their communication cost compared to DDFB.
As a consequence, larger acceleration factors could be expected for these networks.
The scalability of both DnCNN and DRUNet is however mitigated by the larger number of communication phases required (20 and 58) compared to DDFB (8 phases), leading to large runtime overheads.

\subsection{Reconstruction quality experiments}
\label{sec:exp_results:quality}

\begin{table}[t]
    \caption{\mb{\textbf{Reconstruction quality for all problems --}} Best and second best are highlighted in bold and underlined, respectively. \mb{The results are the same for all $B \in \{1, 2, 4\}$.}
    }
    \label{tab:metrics_strong_scaling}
    \centering
    \begin{tabular}{llccc}
        \toprule
        \textbf{Task} & \textbf{Prior} & \textbf{rSNR ($\uparrow$)} & \textbf{PSNR ($\uparrow$)} & \textbf{SSIM ($\uparrow$)} \\
        \midrule
        \multirow{4}{*}{Inpainting}
            & TV & $21.12$ & $27.42$ & $0.68$     \\
            & DDFB ($K=4$) & $\underline{22.88}$ & $\underline{29.19}$ & $\underline{0.79}$   \\
            & DnCNN & $20.76$ & $27.07$ & $0.74$  \\
            & DRUNet & $\mathbf{23.59}$ & $\mathbf{29.90}$ & $\mathbf{0.81}$ \\
        \midrule
        \multirow{4}{*}{\makecell[tl]{Gaussian \\ Deconv.}}
            & TV & $17.56$ & $23.86$ & $0.63$     \\
            & DDFB ($K=4$) & $\mathbf{21.05}$ & $\mathbf{27.35}$ & $\mathbf{0.79}$   \\
            & DnCNN & $18.93$ & $25.23$ & $0.70$  \\
            & DRUNet & $\underline{20.47}$ & $\underline{26.78}$ & $\underline{0.74}$ \\
        \midrule
        \multirow{4}{*}{\makecell[tl]{Poisson \\ Deconv.}}
            & TV & $\underline{19.47}$ & $\underline{25.78}$ & $0.66$     \\
            & DDFB ($K=4$) & $\mathbf{20.31}$ & $\mathbf{26.61}$ & $\mathbf{0.75}$   \\
            & DnCNN & $18.78$ & $25.08$ & $\underline{0.71}$  \\
            & DRUNet & $18.52$ & $24.83$ & $0.61$ \\
        \bottomrule
    \end{tabular}
\end{table}

\begin{figure}[tb]
    \centering
    \setlength{\tabcolsep}{2pt} 
    \renewcommand{\arraystretch}{1}

    \begin{tabular}{vVV}
    \rotatebox{90}{$\overline{\bm{x}}$ and $\bm{y}$}
    &
    \includegraphics[width=0.16\textwidth]{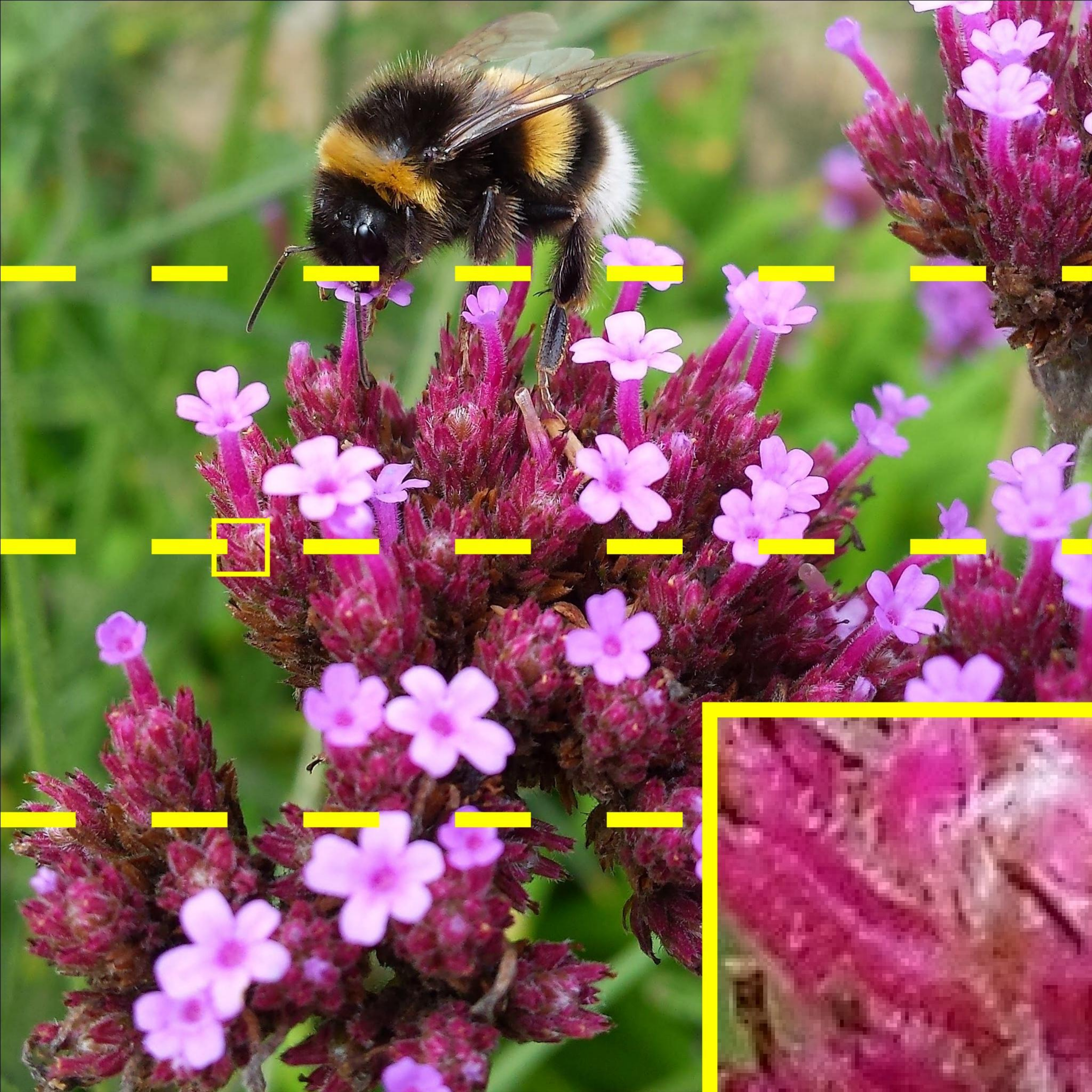} 
    &
    \includegraphics[width=0.16\textwidth]{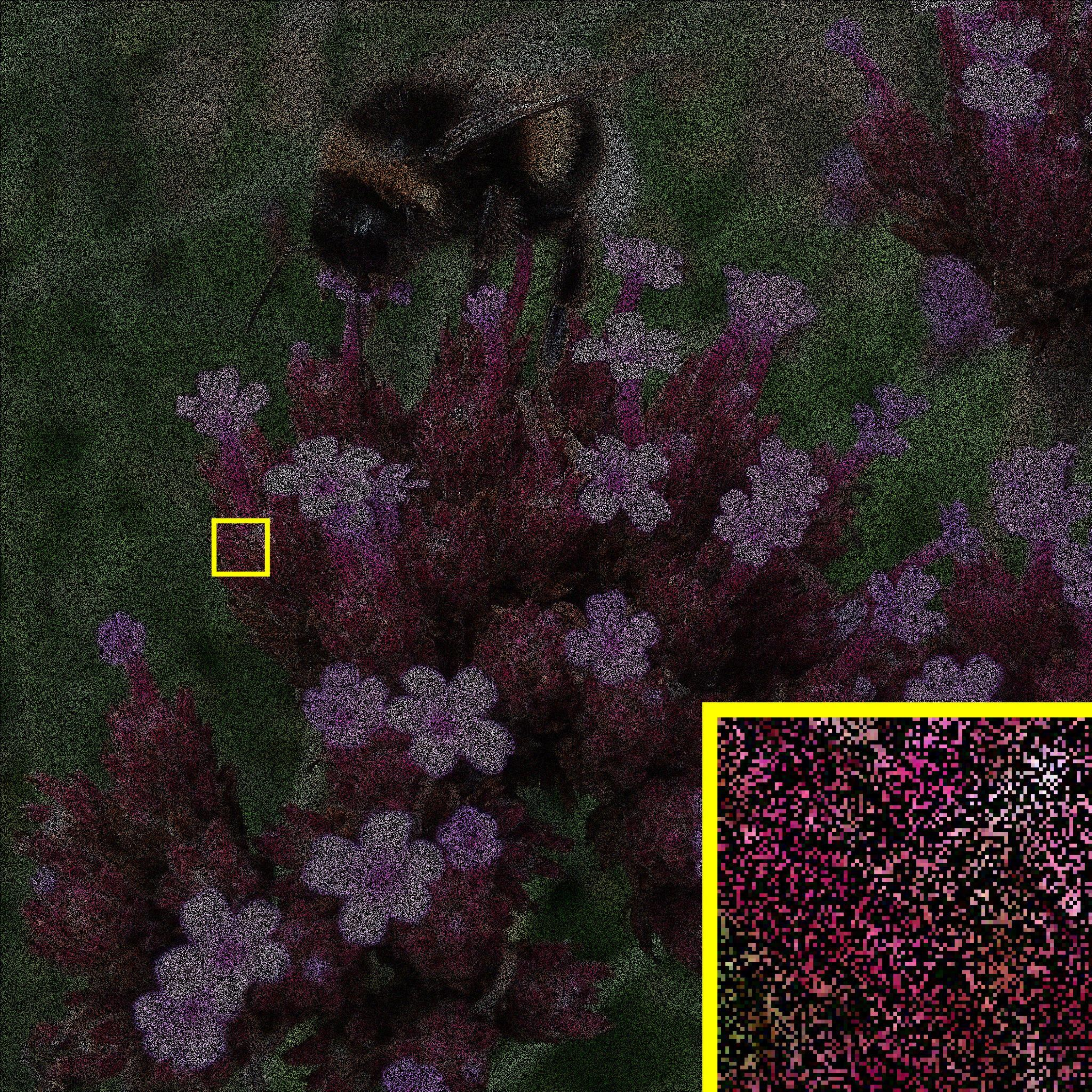} 
    \\
    \rotatebox{90}{TV}
    &
    \includegraphics[width=0.16\textwidth]{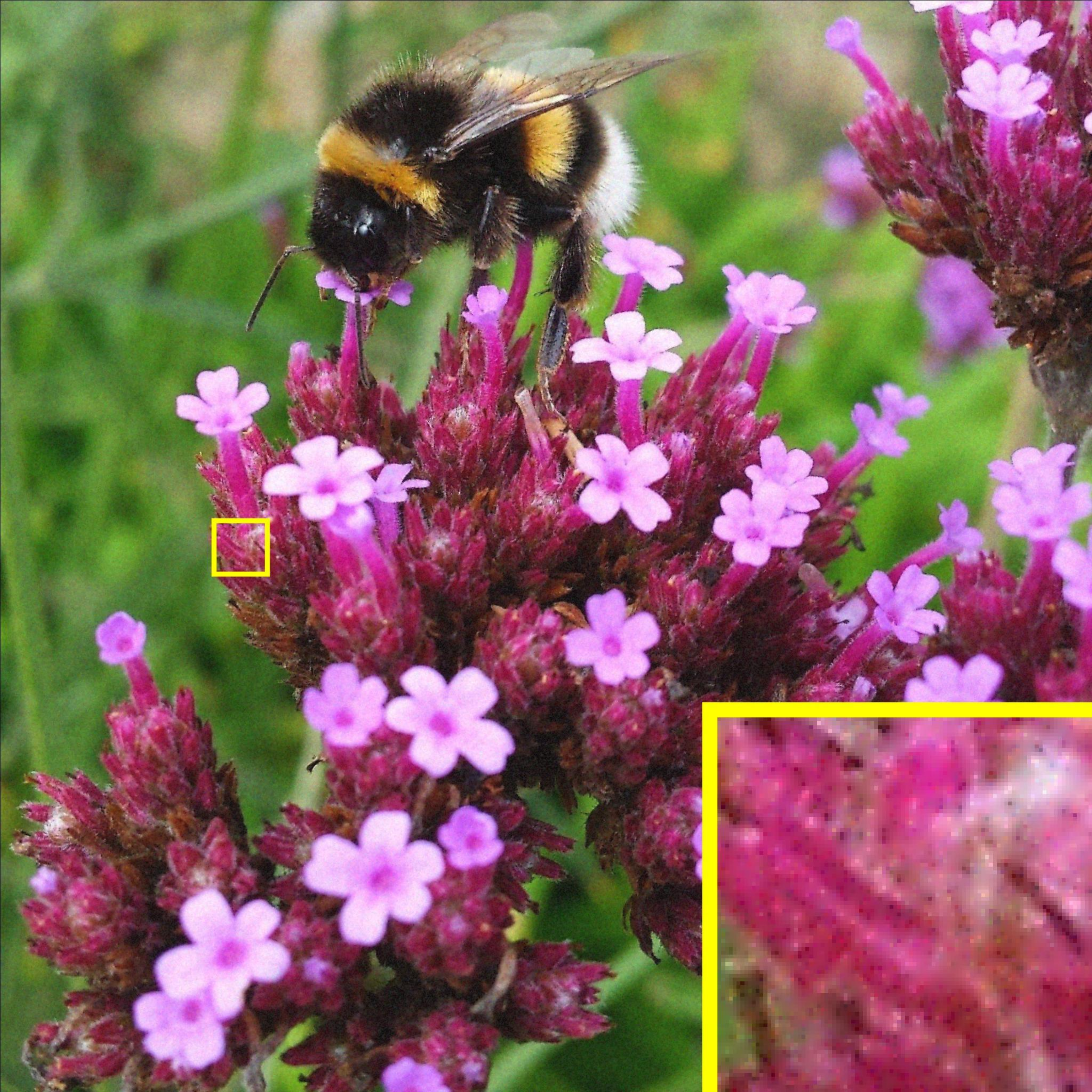} 
    &
    \includegraphics[width=0.16\textwidth]{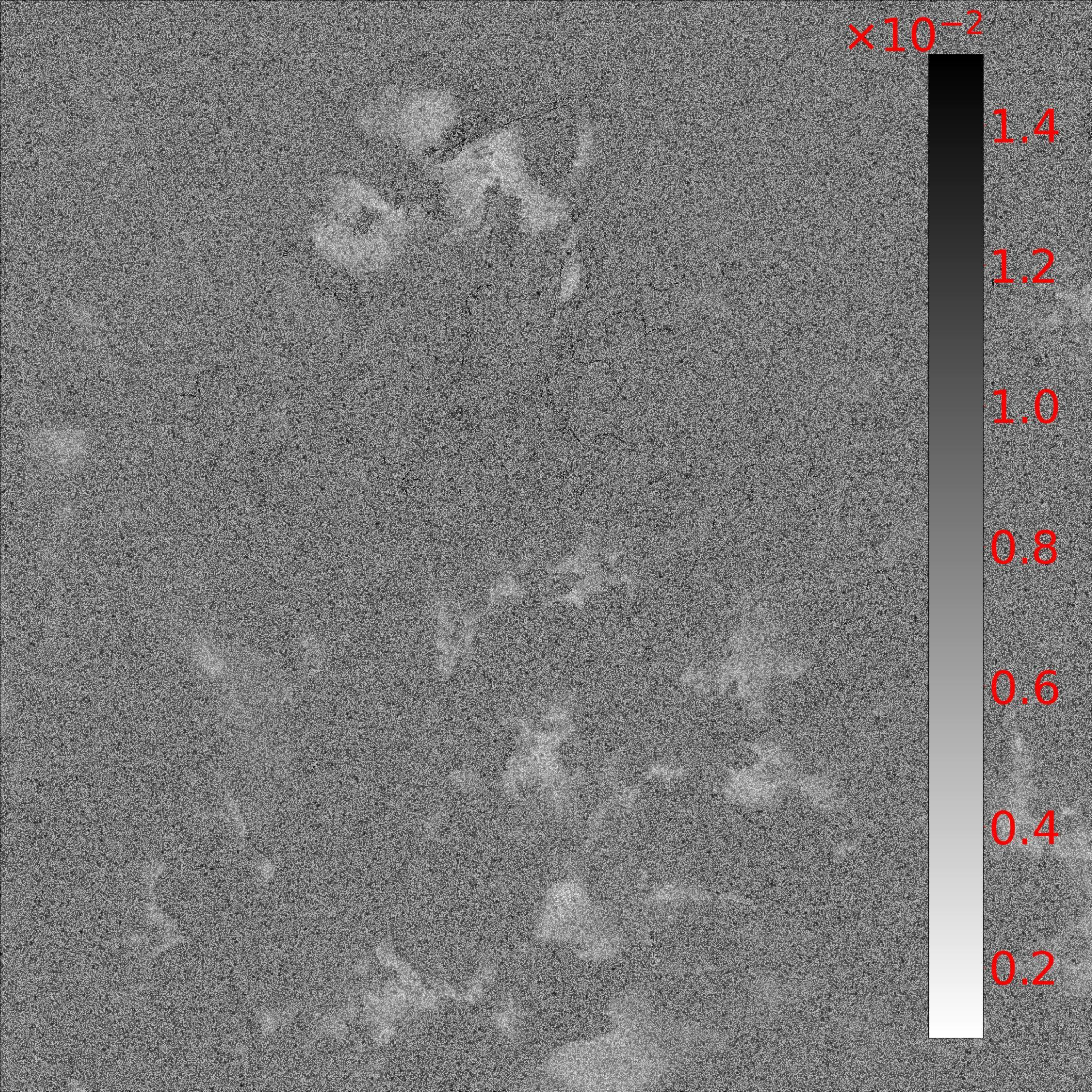} 
    \\
    \rotatebox{90}{DDFB}
    &
    \includegraphics[width=0.16\textwidth]{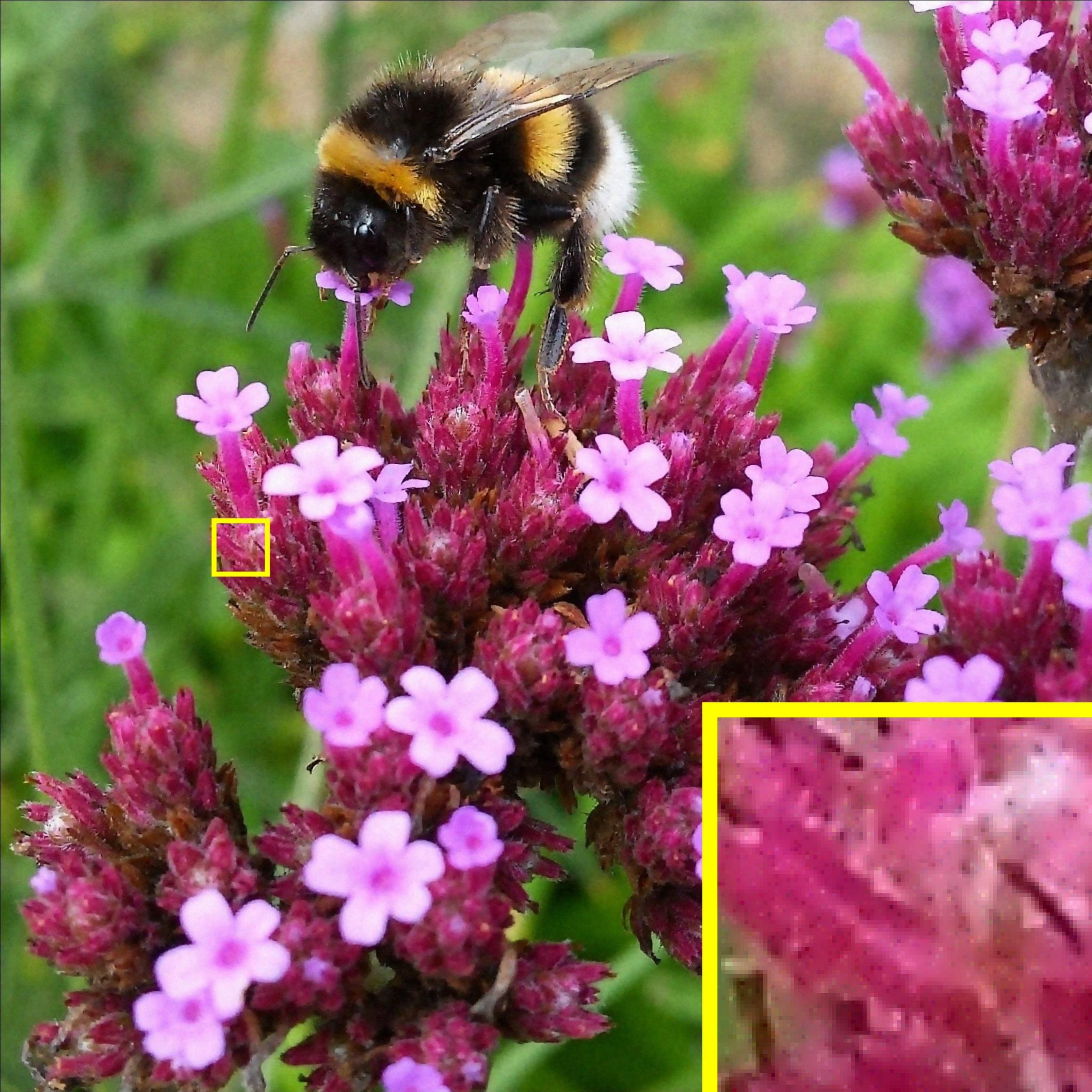} 
    &
    \includegraphics[width=0.16\textwidth]{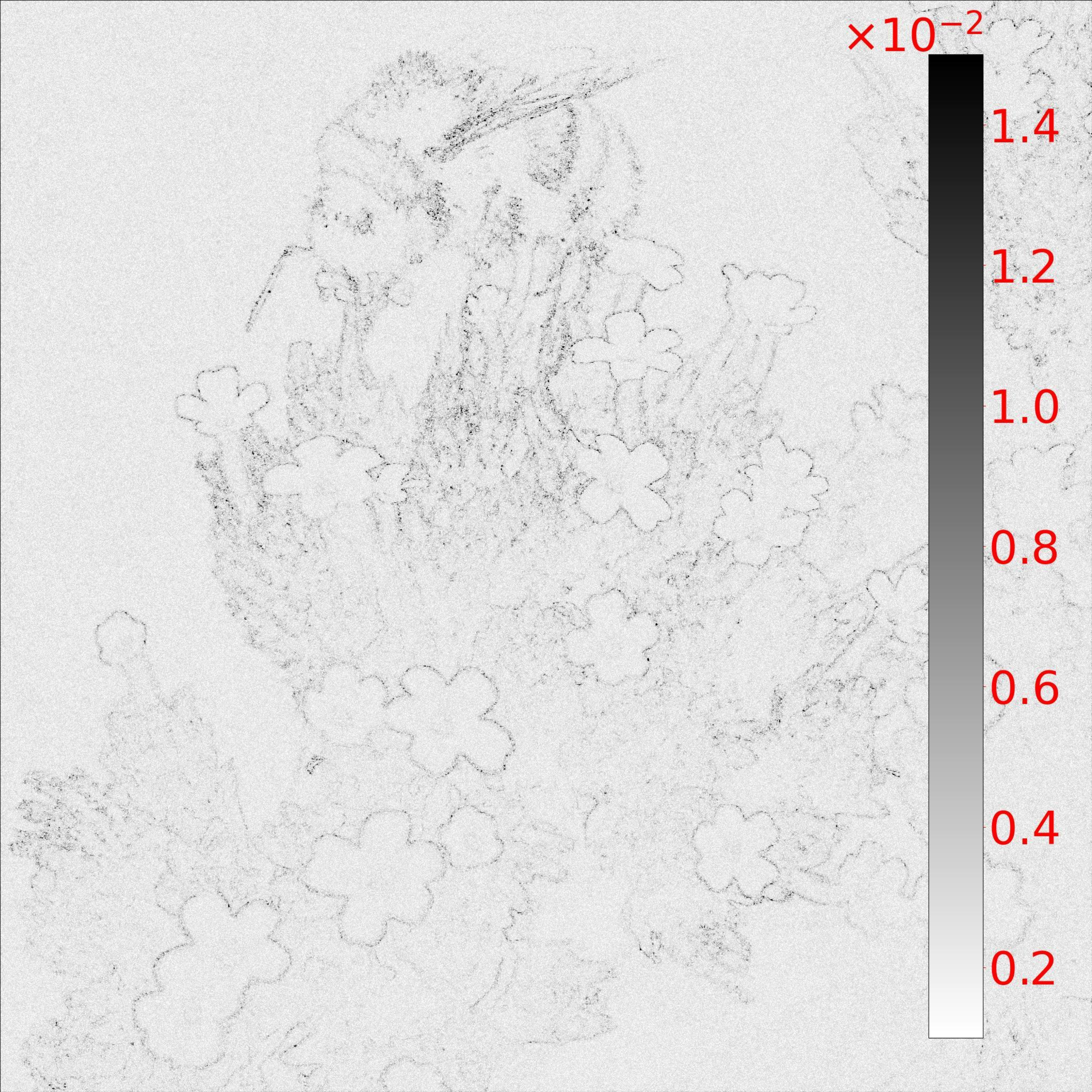} 
    \\
    \rotatebox{90}{DnCNN}
    &
    \includegraphics[width=0.16\textwidth]{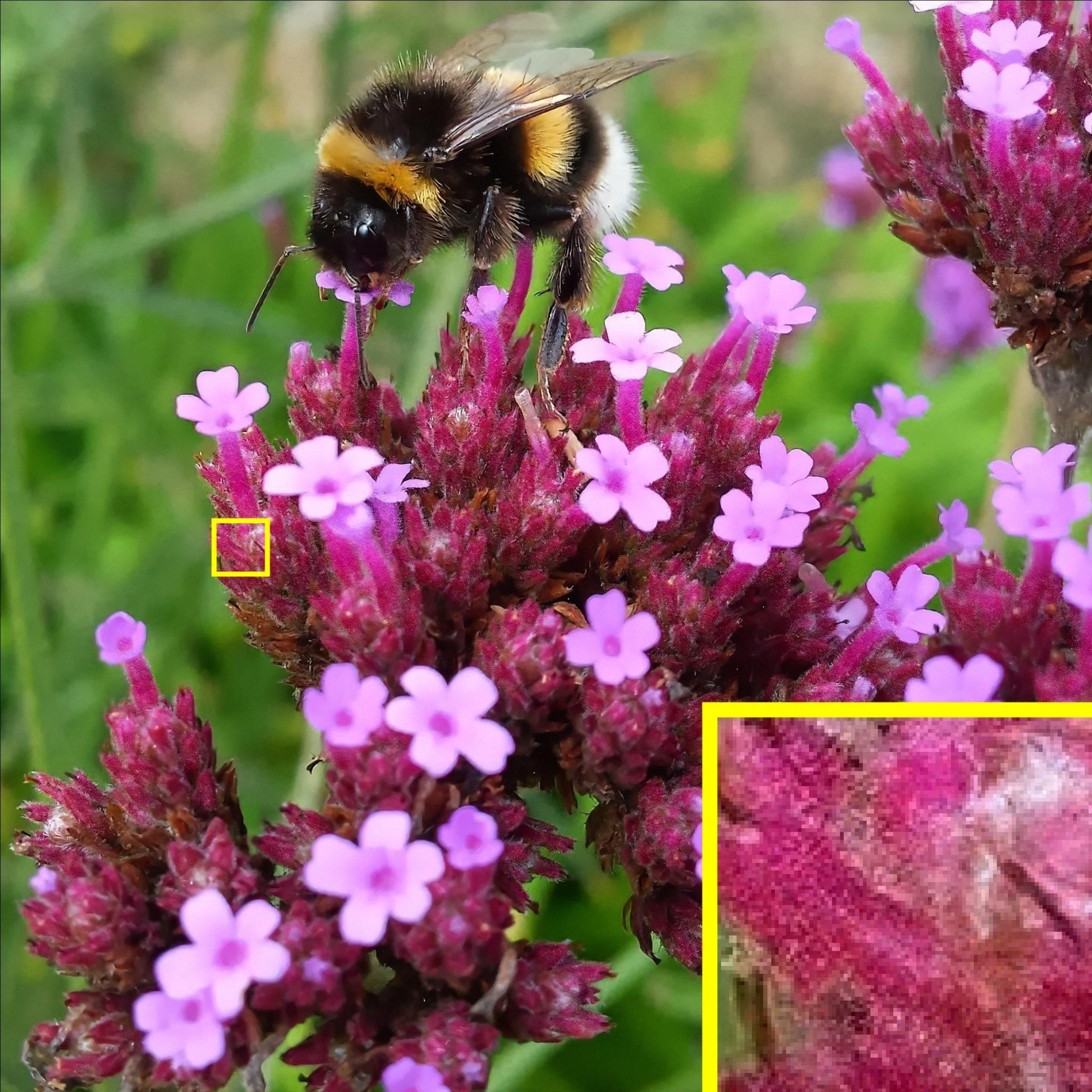} 
    &
    \includegraphics[width=0.16\textwidth]{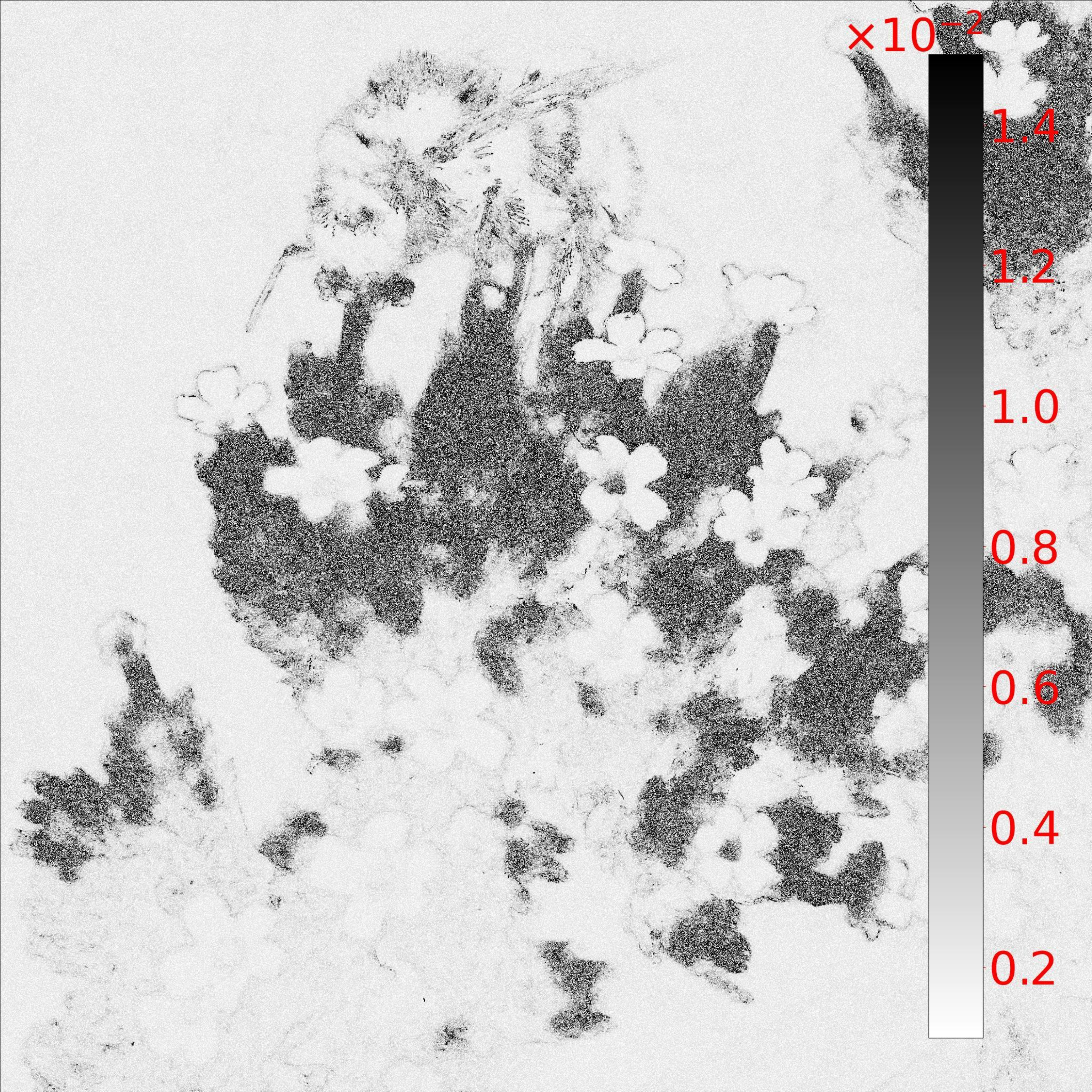} 
    \\
    \rotatebox{90}{DRUNet}
    &
    \includegraphics[width=0.16\textwidth]{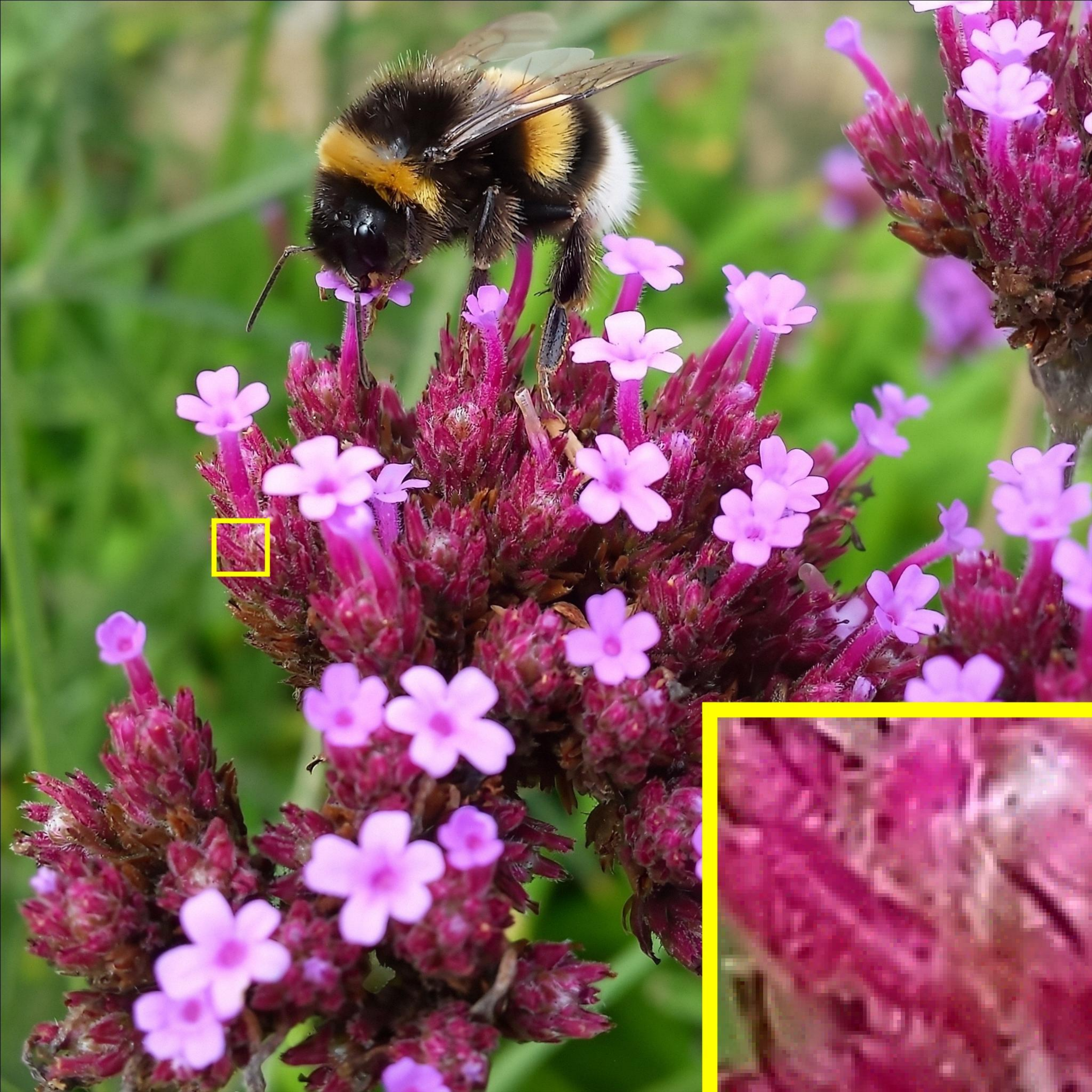} 
    &
    \includegraphics[width=0.16\textwidth]{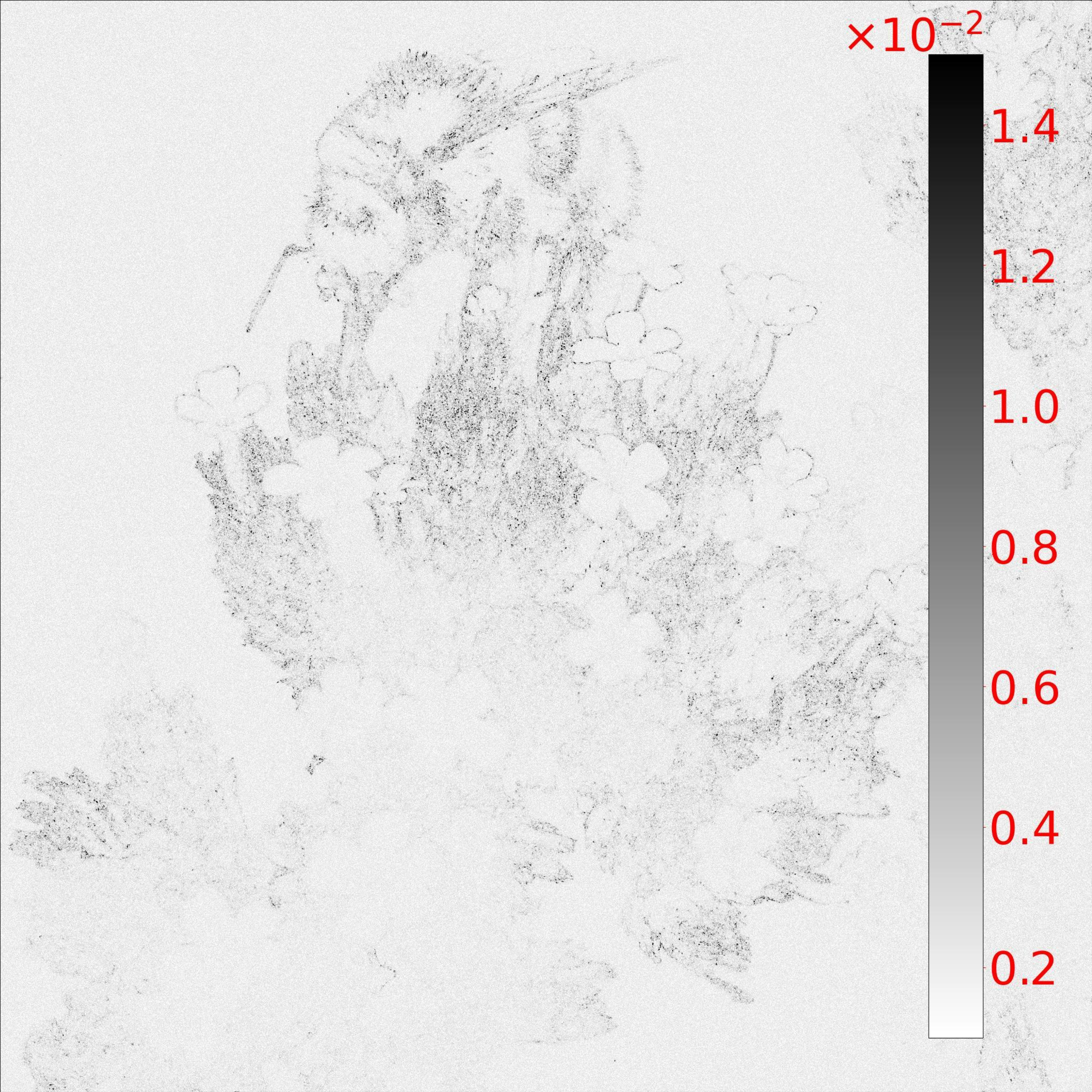} 
    \end{tabular}

    \caption{
        \textbf{Inpainting results --}
        Restoration results with~\Cref{algo:dsgs_pnp_ula_psgla} using $\nblocks = 4$ workers.
        First row: ground truth $\overline{\x}$ and observations $\y$.
        Last 4 rows, top to bottom: MMSE (left) and pixel variance of the red channel (right) using either TV, DDFB, DnCNN, or DRUNet.
        Dashed lines on the ground truth image show the partition considered for $\bm{x}$.
    }
    \label{fig:sequential_comparison}
\end{figure}

The quality of the MMSE estimates obtained with~\Cref{algo:dsgs_pnp_ula_psgla} is summarized in~\Cref{tab:metrics_strong_scaling}. 
Among the priors tested, DDFB consistently gives estimates with satisfactory and stable metrics, outperforming the TV baseline.

Despite superior performance in pure denoising tasks (see~\Cref{tab:denoising_comparison_metrics}), DnCNN yields poorer estimates.
This degradation stems primarily from the denoiser's design: it performs blind denoising, \emph{i.e.}, without explicit regularization parameter $\epsilon$.
While DnCNN was trained across a range of noise levels, the PnP-ULA framework assumes a fixed denoising level $\epsilon$ throughout the sampling process.
The mismatch between the network's implicit behavior and the fixed parameter required by the sampler complicates tuning and may lead to instability, ultimately degrading the reconstruction quality.

DRUNet achieves the best results for the inpainting task, but its performance deteriorates sharply for the deconvolution problem in presence of Poisson noise.
These poor metrics come from spurious details and high-frequency artifacts in pixels where uncertainty in the observations is the highest.
These artifacts suggest instability during sampling, likely caused by an underestimation of $L_\epsilon$.
A more conservative value could stabilize the sampling at the cost of smaller step sizes, reducing the exploration capability of the Markov chain.

For both the inpainting and Gaussian deconvolution tasks, the best results were obtained using the denoising level $\epsilon = \sigma$ for all PnP priors (\emph{i.e.}, the noise level in the observations).
In this configuration, the denoiser effectively aligns with the degradation model.
For deconvolution under Poisson noise, determining an appropriate denoising level is considerably more challenging. 
This mostly impacts the reconstruction performance with DRUNet.

Finally, results for the inpainting problem are shown in~\Cref{fig:sequential_comparison}. 
Restored images for the deconvolution problems can be found in the supplementary material, as well as reconstruction error maps for the estimates.
Pixels with the highest variance also have the highest errors.
The MMSE and pixel-wise variance estimates are displayed for the inpainting task of an $N = 3 \times 2048^2$ image using $B=4$ worker (see~\Cref{sec:gaussian_inpainting}).
Note that \Cref{algo:dsgs_pnp_ula_psgla} produces estimates of the same quality for $\nblocks \in \{1, 2, 4\}$. 
In particular, \mb{deterministic+} operation outputs coincide for any number of workers $\nblocks$ down to machine precision, except the generation of pseudo-random numbers.

\subsection{Scalability experiments}
\label{sec:exp_results:scalability}

Both strong and weak scaling results are influenced by the overall balance between computing and communication costs.
In particular, operators in the likelihood term can significantly affect scalability.
For inpainting, all operations in the likelihood are element-wise and fully local (\emph{i.e.}, no communication is required).
In contrast, the likelihood for deconvolution tasks induce communication costs scaling with the kernel size. 
This distinction has an impact on scalability, especially for tasks based on a prior with a low computation cost, such as the TV.

\subsubsection{Strong scaling}
\label{sec:exp_results:strong_scaling}

\begin{table}[t]
    \caption{\mb{\textbf{Strong scaling experiments --}} 
    Runtime per iteration (in ms) and speed-up for $B\in \{2, 4\}$ with respect to $B=1$.
    Results in red highlight configurations for which there is no practical benefit from a distributed over a serial implementation.
    }
    \label{tab:runtime_strong_scaling}
    \centering
    \begin{tabular}{llrrr}
        \toprule
        \textbf{Task} & \textbf{Prior} & \multicolumn{3}{c}{\textbf{Time in ms (speed-up w.r.t $B = 1$)}} \\
        \cmidrule{3-5}
        & & \multicolumn{1}{c}{$B=1$} & \multicolumn{1}{c}{$B=2$} & \multicolumn{1}{c}{$B=4$} \\
        \midrule
        \multirow{4}{*}{Inpainting}
            & TV & $7.09$ & {\color{red}$7.18~(0.99)$} & $4.17~(1.40)$ \\
            & DDFB & $88.97$ & $65.09~(1.37)$ & $36.80~(2.42)$ \\
            & DnCNN & $342.13$ & $241.40~(1.42)$ & $176.29~(1.94)$ \\
            & DRUNet & $1505.90$ & $998.22~(1.51)$ & $771.46~(1.95)$ \\
        \midrule
        \multirow{4}{*}{\makecell[tl]{Gaussian \\ Deconv.}}
            & TV & $11.71$ & {\color{red}$21.16~(0.55)$} & {\color{red}$24.32~(0.48)$} \\
            & DDFB & $94.84$ & $80.75~(1.17)$ & $57.08~(1.66)$ \\
            & DnCNN & $353.00$ & $251.72~(1.40)$ & $196.05~(1.80)$ \\
            & DRUNet & $1499.61$ & $1001.38~(1.50)$ & $791.41~(1.89)$ \\
        \midrule
        \multirow{4}{*}{\makecell[tl]{Poisson \\ Deconv.}}
            & TV & $14.04$ & {\color{red}$24.24~(0.58)$} & {\color{red}$25.11~(0.56)$} \\
            & DDFB & $97.56$ & $84.05~(1.16)$ & $59.82~(1.63)$ \\
            & DnCNN & $356.75$ & $255.86~(1.39)$ & $197.11~(1.81)$ \\
            & DRUNet & $1514.14$ & $1122.49~(1.35)$ & $793.18~(1.91)$ \\
        \bottomrule
    \end{tabular}
\end{table}

\Cref{tab:runtime_strong_scaling} reports strong scaling results for the applications described in~\cref{sec:applications}. Speed-up values below $1$, highlighted in red, indicate that distributing over multiple workers does not lead to any acceleration (hence do not justify the use of multiple GPUs).
This is the case when using the TV prior: apart from the inpainting experiment with $B=4$, none of the other experiments benefit from the distributed strategy.
Instead, samplers based on deep priors exhibit good scaling behaviour due to a favorable ratio between computation and communication costs, with speed-ups between $1.1$ and $1.5$ for $B = 2$, and $1.63$ and $2.42$ for $B = 4$.

\subsubsection{Weak scaling}
\label{sec:exp_results:weak_scaling}

\Cref{tab:runtime_weak_scaling} presents weak scaling results, obtained by increasing both $B$ and $N_yN_x$ in the same proportion.
Ideal weak scaling corresponds to a constant runtime over the different configurations.
However, in practice, a gradual degradation in efficiency is often expected as $B$ increases, due to increasing communication overheads.

As for strong scaling, samplers based on deep priors yield better efficiency than TV.
With the latter, local computation costs do not significantly outweight communication overheads.
For inpainting, all denoisers maintain acceptable efficiency levels (between $70\%$ and $80\%$), while TV drops to $51\%$ for $B=2$.
For deconvolution problems, communication costs related to the likelihood increase significantly, as the kernel size increases with the image size.
This results in sharp efficiency drops across all priors: below $60\%$ for DnCNN and DDFB for $B=2$, and below $40\%$ for $B=4$.
The TV prior suffers the most, with $13\%$ efficiency with $B=4$. 

Overall, scalability results confirm the interest of the proposed distributed approach when the computing load on each worker is large enough to justify the communication costs incurred, as for other distributed computing approaches~\cite{Eijkhout2023}.

\begin{table}[t]
    \caption{\mb{\textbf{Weak scaling experiments --}} 
    Runtime per iteration (in ms) and efficiency for $(B, N_yN_x) \in \{(1, 2048^2), (2, 2896^2), (4, 4096^2) \}$.
    Efficiency results indicating poor use of hardware resources (below 30\%) are highlighted in red.
    }
    \label{tab:runtime_weak_scaling}
    \centering
    \begin{tabular}{llrrr}
        \toprule
        \textbf{Task} & \textbf{Prior} & \multicolumn{3}{c}{\textbf{Time in ms (efficiency w.r.t. setting $B = 1$)}} \\
        \cmidrule{3-5}
        & & \multicolumn{1}{c}{$(1,2048^2)$} & \multicolumn{1}{c}{$(2,2896^2)$} & \multicolumn{1}{c}{$(4, 4096^2)$} \\
        \midrule
        \multirow{4}{*}{Inpainting} 
            & TV & $7.09 $ & $13.87~(51\%)$ & $14.47~(49\%)$ \\
            & DDFB & $88.97 $ & $127.30~(70\%)$ & $186.37~(47\%)$ \\
            & DnCNN & $342.13 $ & $468.51~(73\%)$ &  $540.66~(63\%)$\\
            & DRUNet & $1505.90 $ & $1891.94~(80\%)$ & $2180.04~(69\%)$ \\
        \midrule
        \multirow{4}{*}{\makecell[tl]{Gaussian \\ Deconv.}}
            & TV & $11.71 $ & {\color{red}$43.86~(27\%)$} & {\color{red}$92.73~(13\%)$} \\
            & DDFB & $94.84 $ & $159.76~(59\%)$ & $268.02~(35\%)$ \\
            & DnCNN & $353.00 $ & $500.53~(71\%)$ & $620.98~(57\%)$ \\
            & DRUNet & $1499.61 $ & $1931.7~(78\%)$ & $2249.29~(67\%)$ \\
        \midrule
        \multirow{4}{*}{\makecell[tl]{Poisson \\ Deconv.}}
            & TV & $14.04 $ & {\color{red}$48.12~(29\%)$} & {\color{red}$98.24~(14\%)$} \\
            & DDFB & $97.56 $ & $173.99~(56\%)$ & $277.35~(35\%)$ \\
            & DnCNN & $356.75 $ & $509.76~(70\%)$ & $628.69~(57\%)$ \\
            & DRUNet & $1514.14 $ & $1958.7~(77\%)$ & $2260.03~(67\%)$ \\
        \bottomrule
    \end{tabular}
\end{table}


\section{Conclusion and perspectives}
\label{sec:conclusions}


\mb{To address high-dimensional inverse problems exceeding the memory of a single computing device}, this work introduces a distributed PnP Split Gibbs sampler implemented on a Single Program Multiple Data architecture.
It builds on PnP-ULA~\cite{Laumont2022} to leverage a prior based on a lightweight CNN denoiser.
The proposed sampler exploits the locality of the operators involved in the algorithm for an efficient implementation on a multi-GPU architecture, numerically equivalent to its serial counterpart. 

Experiments on high-dimensional inpainting and deconvolution problems illustrated the quality of the estimates obtained with the proposed approach, comparing different choices of denoisers.
The proposed distributed implementation exhibits both strong and weak scaling behaviour.
Moreover, it enabled efficient sampling of high-resolution $4096^2$ color images, impossible to address with a serial sampler.

Future work will compare different communication strategies~\cite{Galerne2024}, and extend the approach to imaging problems based on non-local operators, affected by severe load-imbalance in a distributed setting.
High-dimensional problems with data defined on graphs or hypergraphs will also be investigated.


\IEEEtriggeratref{23} 
\bibliographystyle{IEEEtran}
\bibliography{strings_all_ref,references_bibtex}

\end{document}